\documentclass[%
 reprint,
 amsmath,amssymb,
 aps,
 showkeys,
]{revtex4-2}

\usepackage{graphicx}
\usepackage{dcolumn}
\usepackage{bm}
\usepackage[colorlinks=true,linkcolor=blue,citecolor=red]{hyperref}

\usepackage{multirow} 
\usepackage{subfigure} 

\usepackage{tikz}
\usetikzlibrary{positioning}
\usetikzlibrary{decorations.pathreplacing,calc}
\tikzset{
otimes/.style={
  circle,
  inner sep=0pt,
  node contents={$\odot$},
  scale=2.5
}
}

\begin{document}

\preprint{APS/123-QED}

\title{Integrated Photonic Fractional Convolution Accelerator}

\author{Kevin Zelaya}
 \affiliation{Department of Physics, Queens College of the City University of New York, Queens, New York 11367, USA}
 \email{kevin.zelaya@cinvestav.mx}
\author{Mohammad-Ali Miri}%
 \affiliation{Department of Physics, Queens College of the City University of New York, Queens, New York 11367, USA}
 \affiliation{%
Physics Program, The Graduate Center of the City University of New York, New York, New York 10016, USA,
}%
\email{mmirilab@gmail.com}

\date{\today}

\begin{abstract}
An integrated photonic circuit architecture to perform a modified-convolution operation based on the Discrete Fractional Fourier Transform (DFrFT) is introduced. This is accomplished by utilizing two nonuniformly coupled waveguide lattices with equally spaced eigenmode spectra, the lengths of which are chosen so that the DFrFT and its inverse operations are achieved. A programmable modulator array is interlaced so that the required fractional convolution operation is performed. Numerical simulations demonstrate that the proposed architecture can effectively perform smoothing and edge detection tasks even for noisy input signals, which is further verified by electromagnetic wave simulations. Notably, mild lattice defects do not jeopardize the architecture performance, showing its resilience to manufacturing errors.

\end{abstract}

\maketitle


\section{Introduction} 

Convolution is one of the fundamental mathematical operations in signal processing which is core to various filtering tasks \cite{oppenheim1997signals}. By convolving a signal with a filter or kernel, one can identify patterns, detect edges, and remove noise, making it an essential tool in many applications such as image and audio processing. Since the introduction of convolutional neural networks, which have been largely successful in various applications, convolution has gained further interest as one of the most important processes for feature extraction in deep learning \cite{goodfellow2016deep}. In optics, since the early days of optical signal processing, there has been great interest in optically performing the convolution operation \cite{Lugt1964, Weaver:66, davis1989compact, bauchert1998data, lu2005neural, chang2018hybrid, miscuglio2020massively}. In free space optics, high-dimensional convolution is readily performed in the Fourier domain by using the so-called 4f system that utilizes two lenses for realizing Discrete Fourier Transform (DFT), in the inverse operation on spatially modulated light wavefront, while the kernel is encoded using electronically-controllable spatial light modulating devices \cite{goodman2005introduction}. In integrated photonics, convolution has been performed in the form of a matrix-vector multiplication and by utilizing wavelength-division multiplexing to encode information in different colors \cite{feldmann2021parallel, mehrabian2019winograd}. These are promising devices that attract wide attention for their potential to speed up classification problems in artificial neural networks~\cite{zhu2022space,mehrabian2019winograd,colburn2019optical}. Nevertheless, on-chip realization of the convolution operator in the Fourier domain has remained challenging, given the lack of a compact integrated DFT unit. Recently, there have been proposals for performing convolution on-chip by utilizing multimode interference in broad-area waveguides with large footprints to approximate the DFT operation \cite{zhu2022space}. 

The fractional Fourier transform (FrFT) was first introduced by Namias~\cite{Nam80} as a way to extend the conventional Fourier transform (FT) for studying the dynamics of free and forced quadratic systems in quantum mechanics. Since its conception, it has been the subject of extensive research due to its potential applications in areas like signal processing and optics, including optimal filtering~\cite{Kut97,Ozt21} and time-frequency analysis~\cite{Sta03,Shi20}. Despite its usefulness, the FrFT does not straightforwardly inherit all the desired mathematical properties of the FT, namely, the translation invariance, correlation, and convolution of signals. In particular, several definitions of the fractional convolution operation have been introduced, fulfilling different properties. For instance, Mendlovic and Ozaktas~\cite{Men93} define it by replacing the conventional FT with the FrFT of arbitrary order. Generalized convolution and product theorems using FrFT have been reported in the literature~\cite{Alm97,Zay98,Tor10,Shi14}, which are not strictly equivalent among themselves as they are based on modified translation invariance theorems; as a result, the product theorem gets replaced by a more relaxed condition. Likewise, several definitions and constructions of the discrete fractional Fourier transform (DFrFT) have been reported in the literature, each with different properties. Particularly, Ref.~\citep{Ata97} reported a construction based on the set of Krawtchouk polynomials and its application using grated index media. In turn, Candan \textit{et al.}~\cite{Can00} introduce an alternative construction in terms of the Harper equation, and the so-called Jx photonic lattice~\cite{Wei16} provides another mechanism to implement the DFrFT in waveguide arrays on chip-scale devices~\cite{Hon22}. Furthermore, these lattices have been experimentally realized in the microwave domain using interdigital capacitors to generate the required coupling strength between microstrip transmission lines~\cite{keshavarz2023real}.

In this letter, we introduce a modified convolution operation based on the discrete Fractional Fourier transform, implemented in a simple photonic circuit involving only waveguides and modulator arrays. Mainly, the photonic Jx lattice~\cite{Wei16, Tsc18, Hon22} is used as the generator of the DFrFT operation. In contradistinction to the DFT, which relies on bulky lens setups that compromise its scalability, the Jx photonic lattice is reasonably easy to fabricate through open-access foundries and can be built up in a relatively small chip-scale setup. Furthermore, for large input ports, the DFrFT converges to the continuous FT for the appropriate length, and thus any operation involving the DFrFT is expected to converge to that of the FT, including the convolution operation. This establishes a comparison point for the results obtained from our architecture and those expected from a conventional convolution scheme. Electromagnetic wave simulations are carried out to test the performance of the proposed device, and error defects are separately studied to analyze the resilience of the architecture.

\section{Fractional Convolution}
\label{sec:SETUP}

Consider the vectors $\boldsymbol{f}=(f_{-j},\ldots,f_{j})^{T}\in\mathbb{C}^{N}$ and $\boldsymbol{k}=(k_{-j},\ldots,k_{j})^{T}\in\mathbb{C}^{N}$, where $N=2j+1$, $j\in\mathbb{Z}^{+}$, and $\boldsymbol{f}^{T}$ stands for the matrix transposition. The convolution of $\boldsymbol{f}$ and $\boldsymbol{k}$, henceforth considered as the input signal and convolution kernel, respectively, is defined as $(\boldsymbol{f}\ast \boldsymbol{g})_{p}=\sum_{q}f_{q}k_{p-q}$. Alternatively, one can rewrite the latter in terms of the \textit{discrete Fourier transform} (DFT) of the aforementioned vectors, $\mathcal{F}[\boldsymbol{f}]=\boldsymbol{F}$ and $\mathcal{F}[\boldsymbol{k}]=\boldsymbol{K}$. This allows rewriting the convolution operation as $\sqrt{N}\mathcal{F}[(\boldsymbol{f}\ast \boldsymbol{k})]_{p}=\boldsymbol{F}_{p} \odot \boldsymbol{K}_{p}\equiv(F_{-j}K_{-j},\ldots,F_{j}K_{j})^{T}$, with $\odot$ denoting the pointwise vector multiplication or Hadamard product. The DFT is defined by the matrix multiplication $\mathcal{F}[\boldsymbol{f}]=\mathbb{F}\boldsymbol{f}$, with the matrix elements $\mathbb{F}_{p,q}=e^{-2i\pi \frac{pq}{N}}/\sqrt{N}$. The convolution is then recovered by performing the inverse DFT. Although mathematically equivalent, the latter form is more convenient for physical applications as the Fourier transform (FT) can be implemented using a conventional lens configuration~\cite{goodman2005introduction}. 

Throughout this work, we exploit the latter construction in order to implement an analogous setup based on the proper discrete fractional Fourier transform (DFrFT) and its asymptotic properties. Following the continuous fractional convolution proposed in~\cite{Men93} (see also~\cite{Tor10,Shi14}), we define the corresponding discrete counterpart of order $\alpha$ as
\begin{equation}
\label{Conv-1}
\sqrt{N} \left( \boldsymbol{f} \underset{\alpha}{\ast} \boldsymbol{k} \right)_{q}=\mathcal{F}^{2\pi-\alpha}[\boldsymbol{F}(\alpha)\odot\boldsymbol{K}(\alpha)]_{q} , 
\end{equation}
where $\boldsymbol{F}(\alpha):=\mathcal{F}^{\alpha}[\boldsymbol{f}]$ and $\boldsymbol{K}(\alpha):=\mathcal{F}^{\alpha}[\boldsymbol{k}]$ are the \textit{discrete fractional Fourier transform} (DFrFT) of order $\alpha$ of $\boldsymbol{f}$ and $\boldsymbol{k}$, respectively.

Before proceeding with the definition of the DFrFT, it is worth mentioning that the construction in Eq.~\eqref{Conv-1} is supported by the following arguments: (a) The DFrFT used in this work can be manufactured in a compact on-chip device (e.g., in a cost-effective silicon-on-insulator platform) using photonic waveguide arrays, where the fractional order $\alpha$ is associated with the array length. (b) As discussed later, in the limit of large arrays $j\rightarrow\infty$ (also equivalent to $N\rightarrow\infty$), the DFrFT under consideration converges to the continuous FT. Thus any operation involving the latter is recovered, including the continuous convolution.

The Jx photonic lattice is used as the candidate to perform the DFrFT operator in our architecture. The latter is based on the coupled mode theory, which is defined in terms of the symmetric and tridiagonal tight-binding matrix Hamiltonian $\mathbb{H}_{p,q}=\kappa_{p}\delta_{p+1,q}+\kappa_{p-1}\delta_{p-1,q},$ where $\kappa_{p}=\frac{\widetilde{\kappa}}{2}\sqrt{(j-p)(j+p+1)}$ stands for the coupling parameter, $p,q\in\{-j,\ldots,j\}$ the waveguide number (channel), $\widetilde{\kappa}$ a scaling factor, and $N=2j+1$ the total number of waveguides in the array. The eigenvalue problem $\mathbb{H}\boldsymbol{u}^{(n)}=\lambda_{n}\boldsymbol{u}^{(n)}$ is well-known in the context of angular momentum in quantum mechanics~\cite{Nar72}, the eigenmodes and eigenvalues of which are
\begin{equation*}
u_{q}^{(n)}=2^{q}\sqrt{\frac{(j+q)!(j-q)!}{(j+n)!(j-n)!}}P_{j+q}^{(n-q,-n-q)}(0), \quad \lambda_{n}=n,
\end{equation*}
respectively, for $K_{0}=1$, and $q,n\in\{-j,\ldots,j\}$. The upper-index $n$ denotes the lattice eigenmode number, the lower-index $q$ the channel number, and $P_{n}^{(a,b)}(z)$ denotes the Jacobi polynomials~\cite{Olv10}. 

Coupled mode theory dictates the wave evolution of the electric field modal amplitudes $\boldsymbol{E}=(e_1,\cdots,e_N)^T$  through the lattice is ruled by the Schr\"odinger-like equation~\cite{Huang94,Chr03}, $id\boldsymbol{E}/dz=\mathbb{H}\boldsymbol{E}$, where $z$ is the propagation distance (normalized by the inverse of a reference coupling level). That is, the action of the unitary evolution operator $\mathbb{G}(z=\alpha)=e^{-i\mathbb{H}\alpha}$ on a given initial electric field $\boldsymbol{E}(z=0)$ generates the output field $\boldsymbol{E}(z=\alpha)$. Such an evolution operator writes in matrix form as~\cite{Wei13}
\begin{multline}
\label{G-alpha}
\mathbb{G}_{p,q}(\alpha)=\sum_{k} e^{-i\alpha n} u_{k}^{(p)}u_{k}^{(q)}\equiv
i^{p-q}\left[\sin\left(\frac{\alpha}{2}\right)\right]^{q-p} \\
\times\left[\cos\left(\frac{\alpha}{2}\right)\right]^{-q-p} \sqrt{\frac{(j+p)!(j-p)!}{(j+q)!(j-q)!}} \\
\times P_{j+p}^{(q-p,-q-p)}(\cos(\alpha)) ,
\end{multline}
where the unitary operator $\mathbb{G}(\alpha)$ clearly meets all the required properties for the DFrFT. That is (i) unitarity, (ii) additivity rule $(\mathcal{F}^{\alpha}\circ \mathcal{F}^{\beta})[f]=\mathcal{F}^{\alpha+\beta}[f]$, and (iii) the existence of the cyclic index $\mathcal{F}^{\alpha=0}[\cdot]=\mathcal{F}^{2\pi}[\cdot]=\mathbb{I}$~\footnote{This depends on the definition of the DFrFT operator, and the cyclic index identity $\alpha=4$ is also used in the literature.}, with $\mathbb{I}$ the corresponding identity matrix in $\mathbb{C}^{N}$. The cyclic index property $\mathbb{G}(2\pi)\equiv\mathbb{I}$ follows from the equidistant spectrum of $\mathbb{H}$ and the spectral decomposition in Eq.~\eqref{G-alpha}.

Although the latter properties may be considered the essential building blocks to define a DFrFT properly, additional conditions may be taken into account, such as the reduction to the continuous FT for infinitely many input ports~\cite{Ata97,Can00}, which is also the case for the Jx lattice in consideration. It is worth mentioning that the exponential form of the evolution operator is equivalent to the spectral decomposition; nevertheless, the opposite is not necessarily true. Indeed, the DFrFT introduced in Ref.~\cite{Can00} is based on the spectral decomposition using the eigenvectors of a lattice Hamiltonian associated with the Harper equation; however, this is not related to the evolution of the Hamiltonian itself. Such a class of models are challenging for photonic implementations like the one presented in this work. 

For our construction, the inverse order $-\alpha$ is, in practice, implemented for the optical length $2\pi-\alpha$. Furthermore, following the cyclic property, the input signal spreads and reconstructs itself at $\alpha=2M\pi$, for $M\in\mathbb{Z}^{+}$. Lastly, the eigensolutions $\boldsymbol{u}^{(n)}$ converge in the limit $N\rightarrow\infty$ to the continuous Hermite-Gauss modes (see Methods section in~\cite{Wei16}), and thus Eq.~\eqref{G-alpha} converges to the FT for $\alpha=\pi/2$ and $j\rightarrow\infty$. Since our convolution architecture is expected to reduce to the conventional convolution for a large enough number of channels and $\alpha=\pi/2$, one can compare its outcome with that of the conventional convolution.

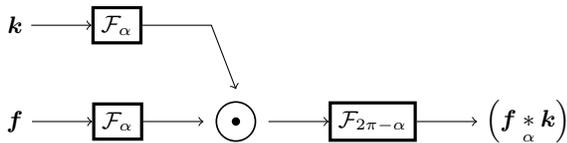
\begin{figure}
\centering
\begin{tikzpicture}[node distance = 8mm]
\node[]      (mainh)                              {$\boldsymbol{k}$};
\node[]        (f)       [below=of mainh] {$\boldsymbol{f}$};
\node[rectangle,draw,very thick]      (Fh)       [right=of mainh] {$\mathcal{F}_{\alpha}$};
\node[]      (empty)       [right=of Fh] {};
\node[circle]      (E1)       [right=of Fh] {};
\node[rectangle,draw,very thick]        (Ff)       [right=of f] {$\mathcal{F}_{\alpha}$};
\node [otimes,right=of Ff,name=cross1];
\node[rectangle,draw,very thick]        (Finv)       [right=of cross1] {$\mathcal{F}_{2\pi-\alpha}$};
\node[]        (conv)       [right=of Finv] {$\left( \boldsymbol{f}\underset{\alpha}{\ast} \boldsymbol{k} \right)$};
\draw[->] (mainh.east) -- (Fh.west);
\draw[->] (f.east) -- (Ff.west);
\draw[->] (Ff.east) -- (cross1.west);
\draw[->] (cross1.east) -- (Finv.west);
\draw[->] (Finv.east) -- (conv.west);
\draw[->] (Fh.east) -- (empty.center) -- (cross1.north);
\end{tikzpicture}
\caption{A block diagram representation of the proposed discrete fractional convolution operation defined in Eq.~\eqref{Conv-1}, where the DFrFT operator is described by the unitary operator Eq.~\eqref{G-alpha}. The symbol $\odot$ represents the pointwise or Hadamard product.}
\label{fig:scheme}
\end{figure}

\begin{figure*}
    \centering
    \includegraphics[width=0.8\textwidth]{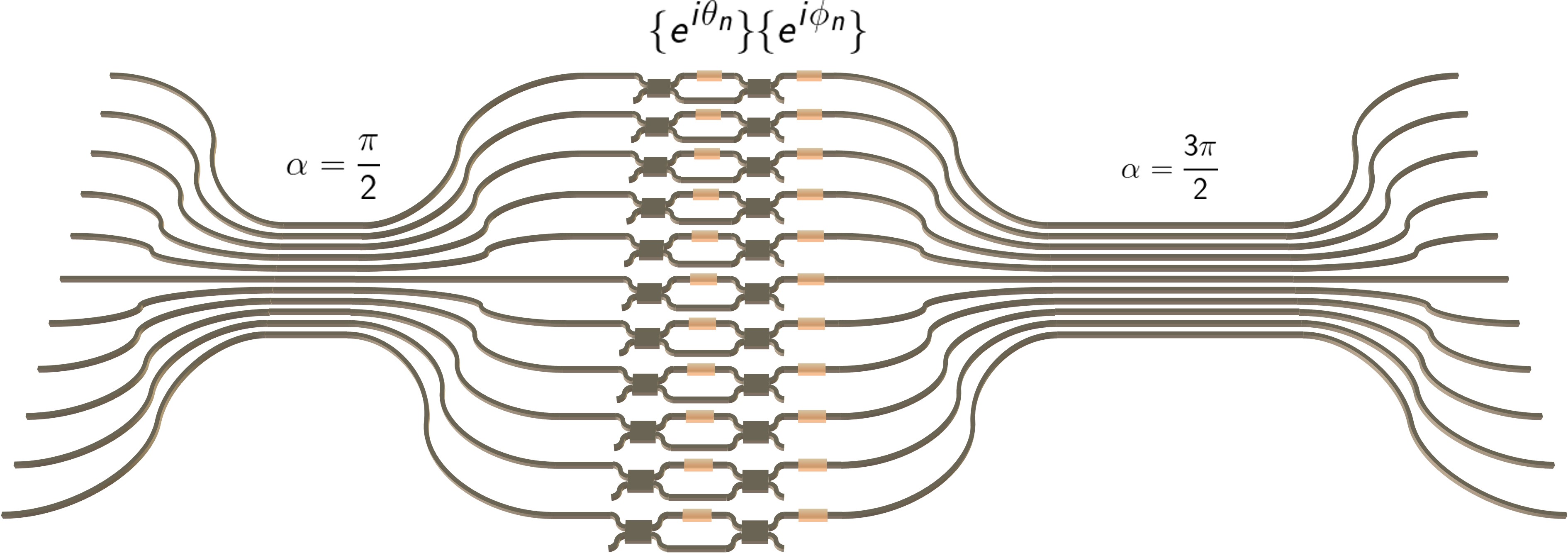}
    \caption{The proposed photonic integrated circuit for implementation of the DFrFT-based convolution accelerator architecture. This architecture is composed of two photonic DFrFT lattices of normalized lengths $\pi/2$ and $3\pi/2$. The intermediate pointwise multiplication is performed through an array of programmable modulators controlled by the set of phase shifters $\{e^{i\theta_{n}}\}$ and $\{e^{i\phi_{n}}\}$.}
    \label{fig:device}
\end{figure*}

\section{Photonic  Implementation}
\label{sec:SIM}
The architecture under consideration is devised by using two photonic Jx lattices, one of (normalized) length $\pi/2$ that performs the DFrFT of the input signal $\boldsymbol{f}$ with fractional order $\alpha=\pi/2$, and another one of length $3\pi/2$ that performs the inverse operation and leads to the output signal. The pointwise product is implemented through a programmable array of Mach-Zehnder interferometers (MZI). Each element of the array is composed of two $50:50$ couplers and two phase shifters $\phi_{n}$ and $\theta_{n}$, with $n=-j,\ldots j$, that enable both amplitude and phase modulation. The field modal amplitude at the $q$-th channel, $e_{q}^{(in)}$, transforms accordingly as $e_{q}^{(out)}=e^{i(\phi_{q}+\theta_{q}/2+\pi/2)}\sin(\theta_{q}/2)e_{q}^{(in)}$. Here, we have used $(\sigma_{0}-i\sigma_{1})/\sqrt{2}$ as the definition for the 50:50 MZI, with $\sigma_{j}$ the conventional Pauli matrices. In this form, $\theta_{q}$ modulates the amplitude at the $q$-th output of the first DFrDT while $\phi_{q}$ modulates the corresponding phase. This allows performing the pointwise multiplication that produces the DFrFT of the convolution kernel instead of the kernel itself. Although one has to electrically adjust the modulator array to generate the DFrFT of the kernel, the latter can be fixed once to perform the convolution for different optical signals fed to the device input channels. 

We now consider our architecture and test it under scenarios in which the well-known convolution is known to work. In this form, we establish a benchmark of the expected convolution results and the actual device output. The operations to be performed and tested throughout the rest of the manuscript are signal displacement, Gaussian filtering, and edge detection. Each of these cases is associated with a specific convolution kernel, and the test is performed with different signals, some including background noise to push forward the capabilities of the architecture.

\subsection{Displacement filter} 
In this particular setup, to simplify and understand the basic properties of our device, we consider the lattice eigenmodes $\boldsymbol{u}^{(n)}$ as the signal input. These eigenmodes can be experimentally produced at the output of an additional photonic Jx lattice of length $\alpha=\pi/2$ by exciting its $n$-th input channel. For instance, one generates a \textit{Gaussian-like} distribution by exciting the $j$-th channel, which is the fundamental mode and, from the oscillation theorem of Hermitian operators~\cite{Nik88}, is the only mode with no oscillations\footnote{Oscillations in the discrete case are defined by their zero crossings. That is, a discrete function $\boldsymbol{f}$ has a zero-crossing at $n$ if $f_{n}f_{n-1}<=0$.}. This is the only mode with such behavior so that, in the continuous limit, it converges to a Gaussian distribution. 

For the convolution kernel, we use a delta-like function $\boldsymbol{k}^{(\delta,r)}_{q}=\delta_{q,r}$, with $r\in\{-j,\ldots j\}$ and $\delta_{q,r}$ the \textit{Kronecker delta function}. This is equivalent to excite only the $r$-th convolution channel of the device. The straightforwards calculation shows that $\boldsymbol{K}^{(\delta,r)}_{p}(\pi/2)=(-i)^{r-p}u^{(m)}_{p}$, whereas the corresponding DFrDT of the Gaussian-like input becomes $\boldsymbol{F}_{p}(\pi/2)=i^{j}u_{p}^{(j)}$. The fractional convolution takes the form $\left(\boldsymbol{f}\underset{\pi/2}{\ast}\boldsymbol{k}\right)_{q}=(-1)^{r}i^{q+r+j}[ u_{0}^{(q)}u_{0}^{(j)}u_{0}^{(r)}+\sum_{p=1}^{j}u_{p}^{(j)}(u_{p}^{(q)}u_{p}^{(r)}+u_{-p}^{(q)}u_{-p}^{(r)}) ]$ (see Appendix~\ref{sec:APPA} for a detailed proof). For even $j$ and following the parity symmetry of the Jx lattice eigenmodes, the convolution vanishes at the odd output channel if $r$ is even; likewise, the output vanishes at the even channel for odd $r$. Furthermore, the roles are interchanged for odd $j$. For the sake of simplicity and without loss of generality, we henceforth focus only on even $j$. To avoid the latter partial suppression at the output, we can exploit the linearity of the convolution scheme Eq.~\eqref{Conv-1} and alternatively use the convolution kernel $\boldsymbol{k}^{(\delta_{2},r)}_{q}=\delta_{q,r}+\delta_{q,r+1}$ composed of two consecutive delta-like signals so that both even and odd channels of $\boldsymbol{u}^{(j)}$ contribute to the convolution process. 

Further remarks can be obtained for more complex input functions. Since the set of eigenfunction $\{\boldsymbol{u}^{(n)}\}_{n=-j}^{j}$ form an orthonormal basis in $\mathbb{C}^{2j+1}$, any arbitrary input signal can be decomposed into the linear combination $\boldsymbol{f}=\sum_{m=-j}^{j}\mathcal{C}_{n}\boldsymbol{u}^{(m)}$, with $\mathcal{C}_{m}=[\boldsymbol{u}^{(m)}]^{\dagger}\boldsymbol{f}$ the corresponding expansion coefficients. For linear combinations of even ($\boldsymbol{f}_{-q}=\boldsymbol{f}_{q}$) or odd signals ($\boldsymbol{f}_{-q}=-\boldsymbol{f}_{q}$), the delta-like convolution kernel produces an output exclusively on the even and odd channels, respectively. 

\begin{figure}
    \centering
    \subfigure[$j=20$]{\includegraphics[width=0.45\textwidth]{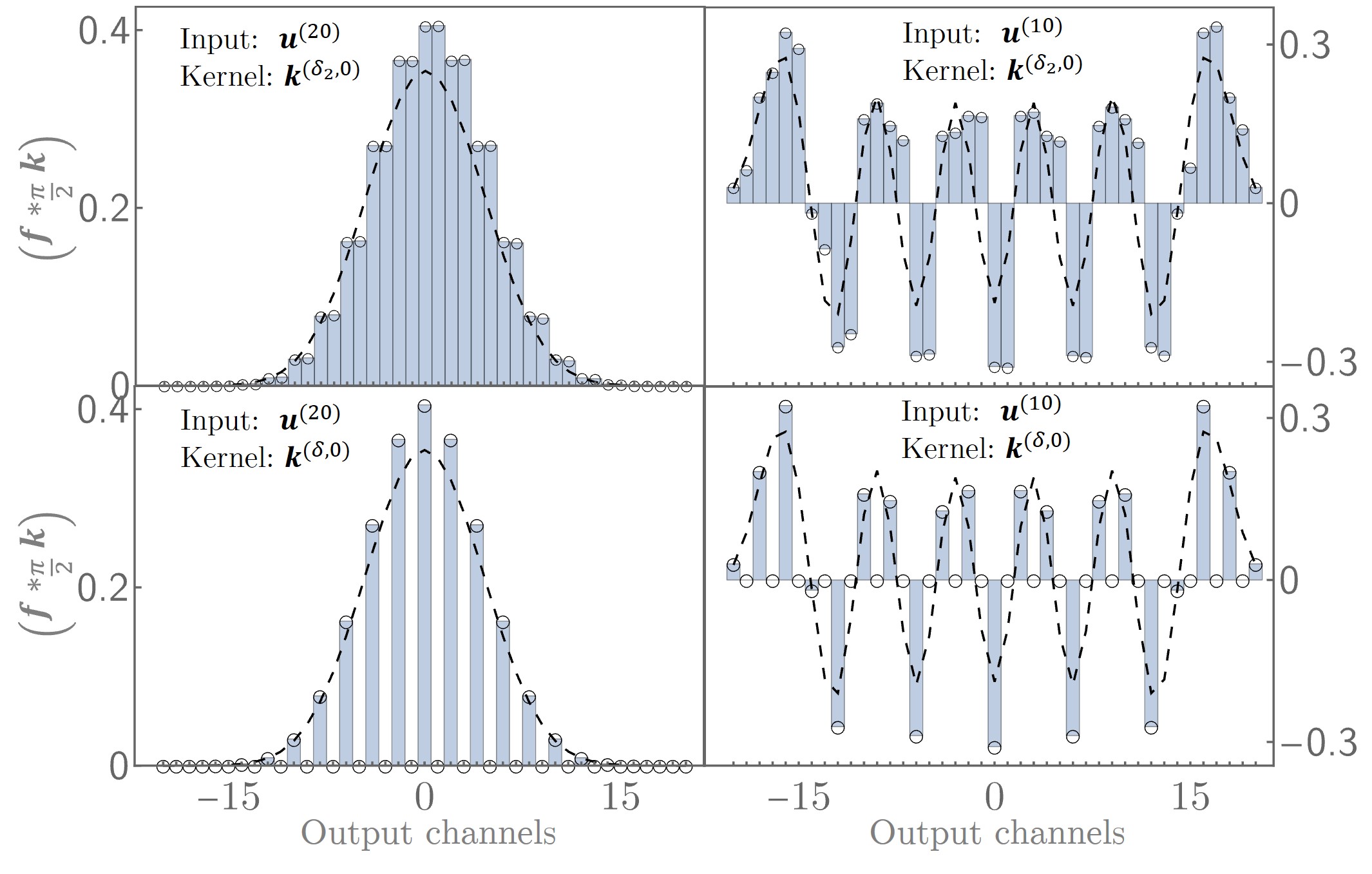}
    \label{fig:filter-1-a}}
    \\
    \subfigure[$j=100$]{\includegraphics[width=0.45\textwidth]{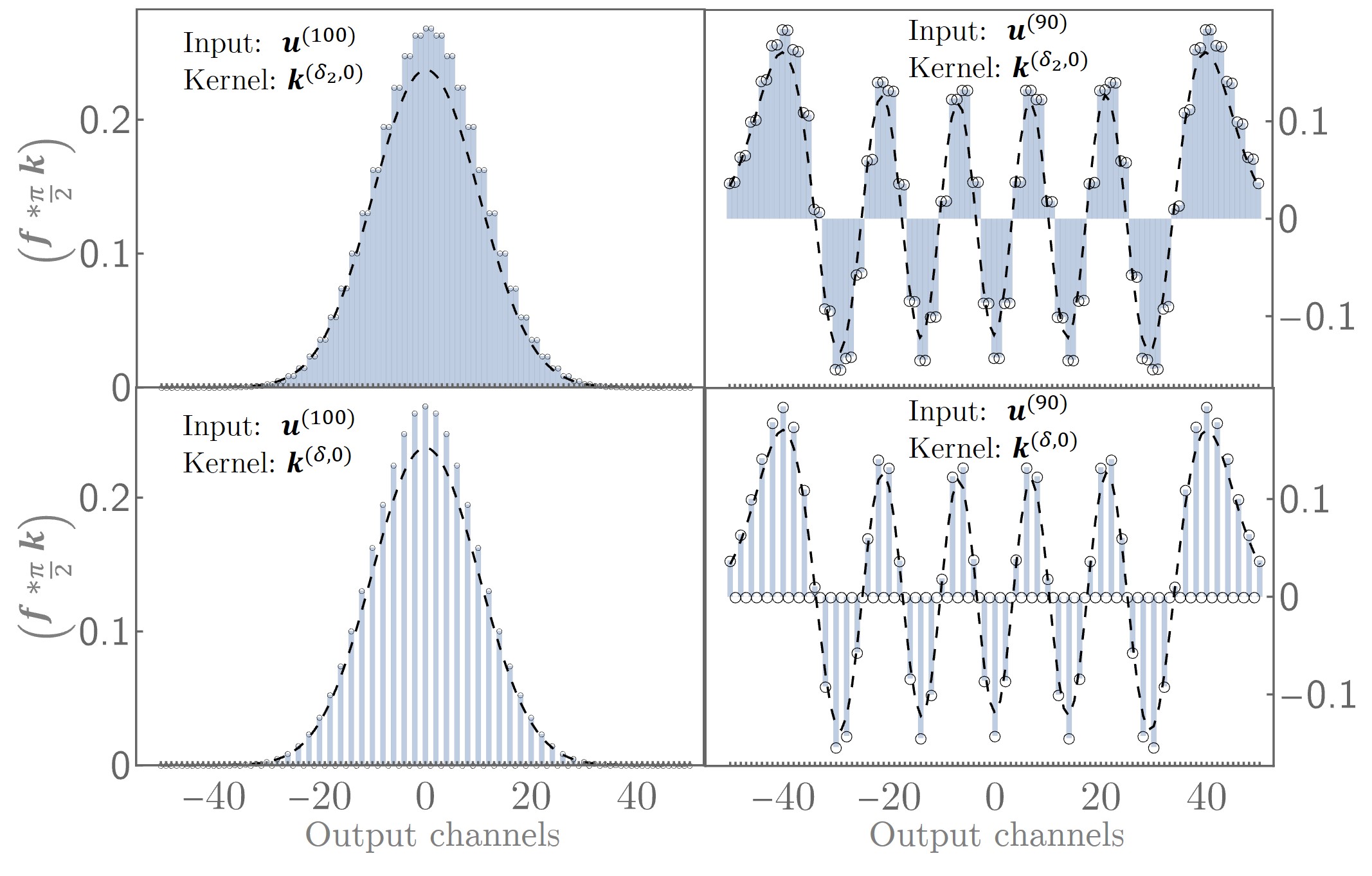}
    \label{fig:filter-1-b}}
    \caption{
    (a) Convolution of the eigenmodes $\boldsymbol{u}^{(20)}$ (left column) and $\boldsymbol{u}^{(10)}$ (right column) with the convolution kernel $\boldsymbol{k}^{(\delta_2,0)}$ (upper row) and $\boldsymbol{k}^{(\delta,0)}$ (lower row) for $j=20$ ($N=41$). (b) Convolution of the eigenmodes $\boldsymbol{u}^{(100)}$ (left column) and $\boldsymbol{u}^{(90)}$ (right column) with the convolution kernel $\boldsymbol{k}^{(\delta_2,0)}$ (upper row) and $\boldsymbol{k}^{(\delta,0)}$ (lower row) for $j=100$ ($N=201$).}
    \label{fig:filter-1}
\end{figure}

It is straightforward to note that the delta-like kernel $\boldsymbol{k}^{(\delta,0)}$ in the conventional convolution (using the DFT) leaves the input signal unchanged, whereas the kernel $\boldsymbol{k}^{(\delta,r)}$ displaces the signal by $r$ units. This establishes a reference framework for our fractional convolution. To this end, let us consider $r=0$ for the delta-like and double-delta kernel, combined with one of the eigenmodes $\boldsymbol{u}^{(s)}$ at the input. The eigenmode number $s$ is been fixed as $s=20$ and $s=10$ for $j=20$ so that we have nodeless and highly oscillatory signals at the input. For $j=100$, we use the corresponding modes, $s=100,90$, that match the number of nodes of the previous case. The resulting fractional convolution is depicted in Figs.~\ref{fig:filter-1-a}-\ref{fig:filter-1-b} for $j=20$ and $j=100$, respectively, where the suppression at the odd channel of the output is evident for the delta-like kernel, as predicted. In turn, the double-delta kernel $\boldsymbol{k}^{(\delta_{2},0)}$ fixes the output suppression and leads to an output signal akin to the respective input eigenmode. To quantify any difference between the input and output, we employ the distance function 
\begin{equation}
\label{distance}
d_{y_{1},y_{2}}=\Vert \boldsymbol{y}_{1}-\boldsymbol{y}_{2} \Vert
\end{equation}
with $\Vert \cdot\Vert$ the Euclidean norm in $\mathbb{C}^{2j+1}$. 

\begin{figure}
    \centering
    \includegraphics[width=0.4\textwidth]{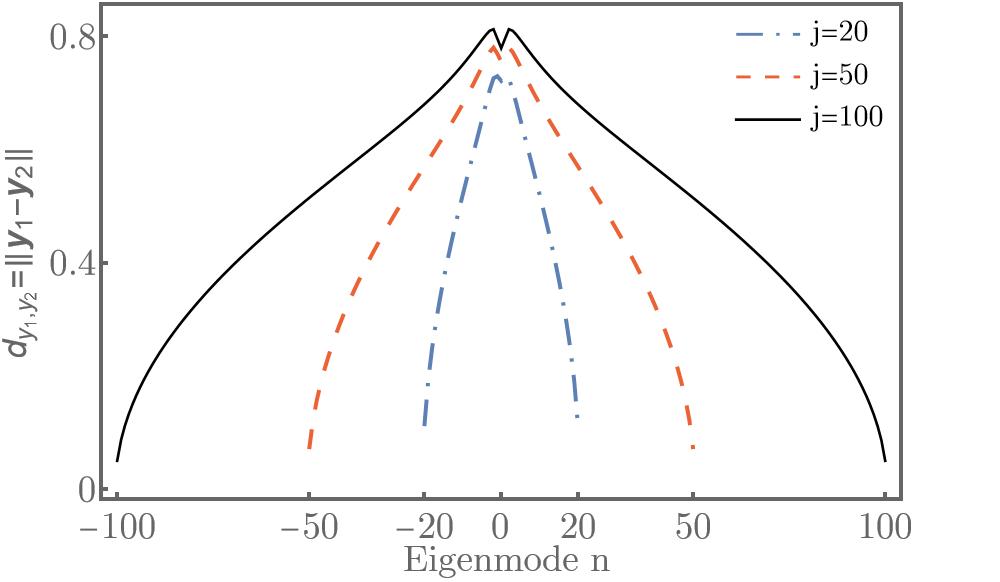}
    \caption{Continuous interpolation of the distance function $d_{y_1,y_2}=\Vert \boldsymbol{y}_{1}-\boldsymbol{y}_{2} \Vert$~(\ref{distance}) between the input eigenmodes $\boldsymbol{y}_{1}=\boldsymbol{u}^{(n)}$ and the corresponding normalized convolution operation using the kernel $\boldsymbol{y}_{2}=\boldsymbol{k}^{(\delta_{2},0)}$. The distance has been plotted as a function of the eigenmode number $n\in\{-j,\ldots,0,\ldots,j\}$ for $j=20,50,100$.
    }
    \label{fig:distance}
\end{figure}

Here we consider $\boldsymbol{y}_{1}=\boldsymbol{u}^{(n)}$ as the input eigenmode and $\boldsymbol{y}_{2}$ the corresponding normalized convolution at the architecture output when the double-delta kernel $\boldsymbol{k}^{(\delta_{2},0)}$ is used. It is worth stressing that the convolution output has been normalized in order to obtain a meaningful comparison. The corresponding results are shown in Fig.~\ref{fig:distance}. On the one hand, it is noted that the distance is smaller for $j=100$ as compared to $j=20$ for the lowest and highest eigenmodes $n=j$ and $n=-j$, leading to a good fidelity at the output. This agrees with the corresponding convolution results presented in Figs~\ref{fig:filter-1}. On the other hand, the distance increases for eigenmodes at and around $n=0$, where more deviations are expected for $j=100$ compared to $j=20$. 

\begin{figure}
    \centering
    \subfigure[$j=20$]{\includegraphics[width=0.45\textwidth]{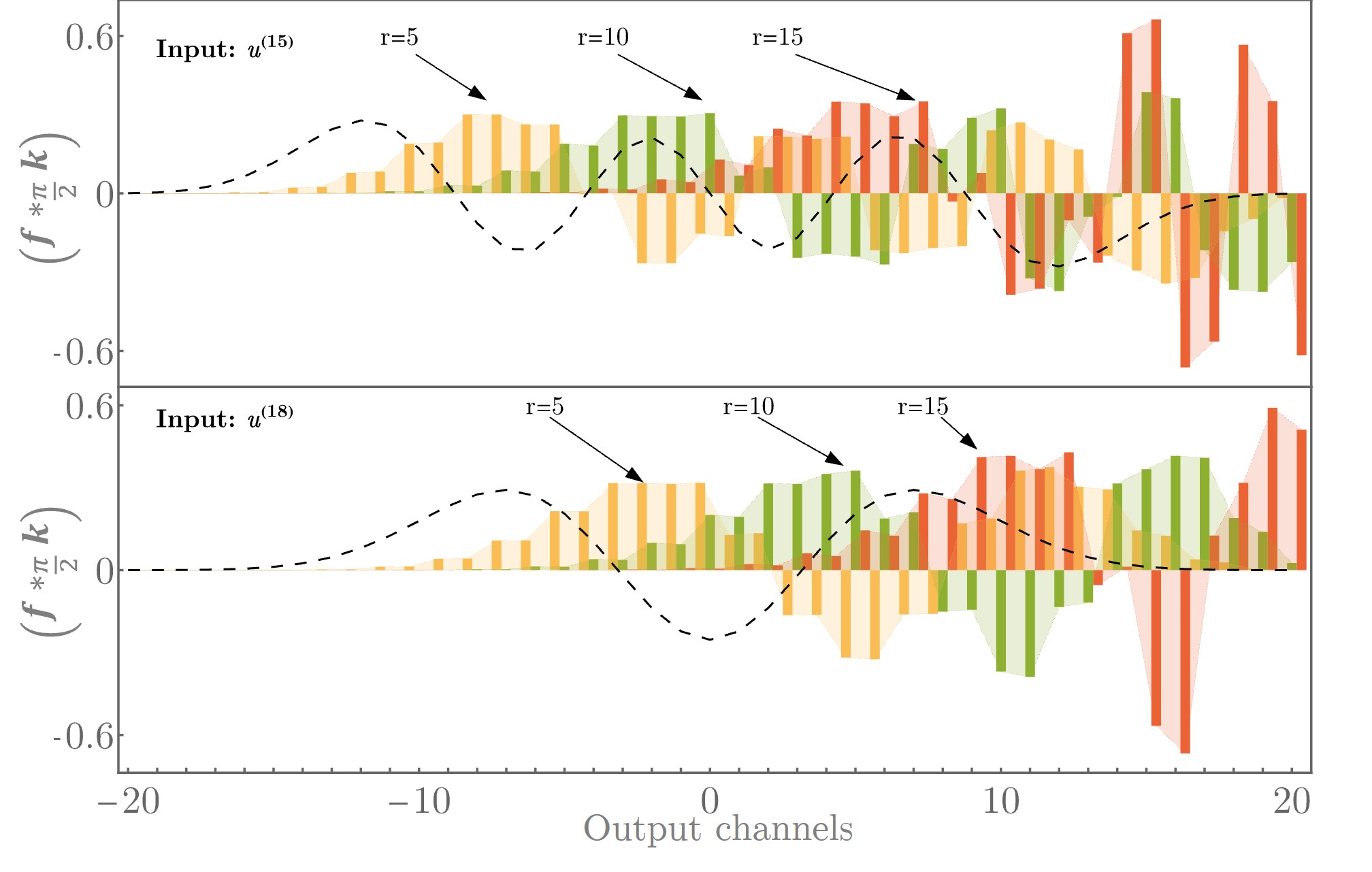}
    \label{fig:filter-2-a}}
    \\
    \subfigure[$j=100$]{\includegraphics[width=0.45\textwidth]{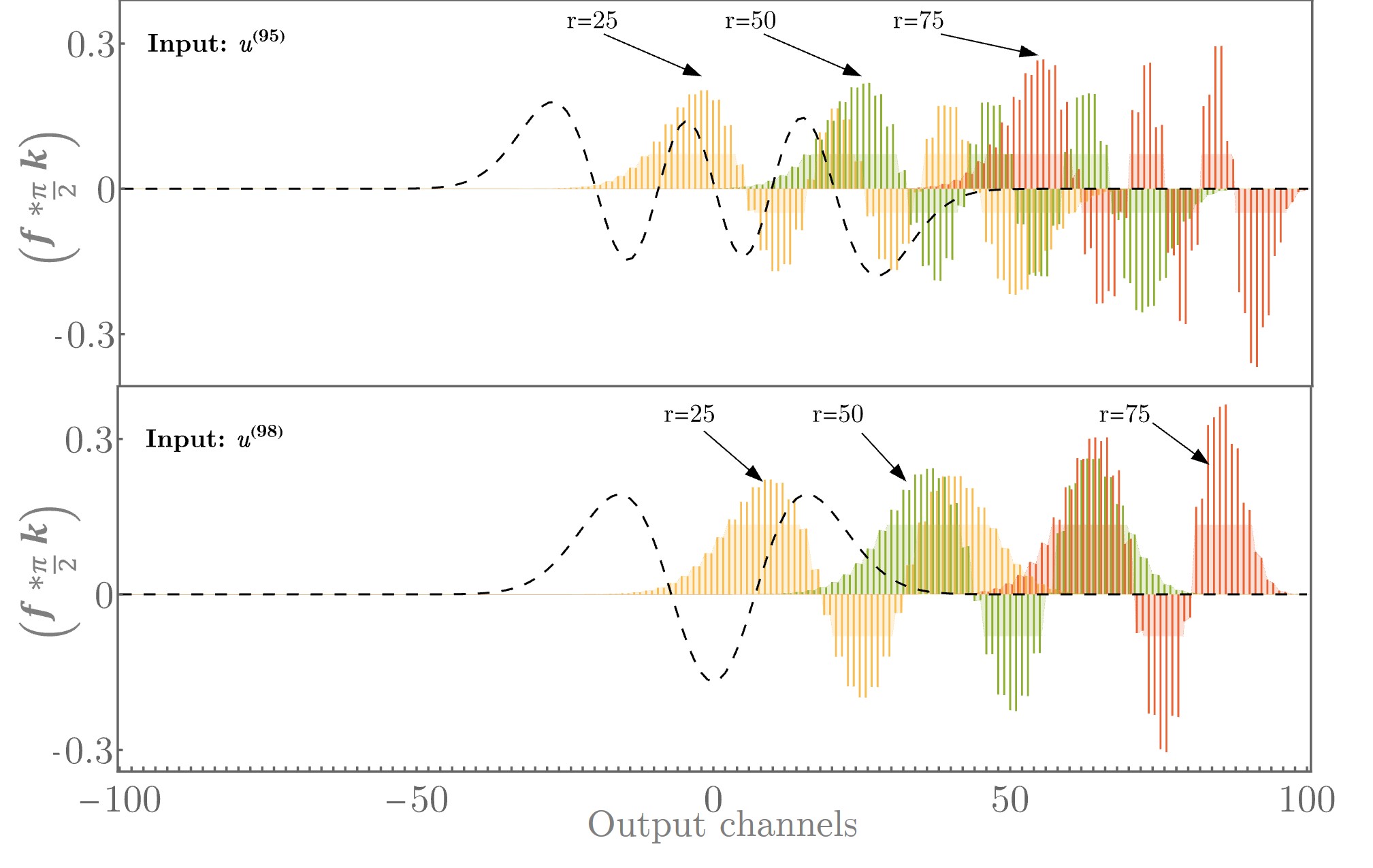}
    \label{fig:filter-2-b}}
    \caption{
    (a) Convolution of the eigenmodes $\boldsymbol{u}^{(15)}$ (upper row) and $\boldsymbol{u}^{(18)}$ (lower row) with the kernel $\boldsymbol{k}^{(\delta_{2},0)}$ for $r=5$ (yellow), $r=10$ (green), and $r=15$ (red) for $j=20$. (b) Convolution of the eigenmodes $\boldsymbol{u}^{(95)}$ and $\boldsymbol{u}^{(98)}$ with the kernel $\boldsymbol{k}^{(\delta_{2},0)}$ for $r=25$ (yellow), $r=50$ (green), and $r=75$ (red) for $j=100$.
    }
    \label{fig:filter-2}
\end{figure}

Once the fidelity of the convolution process with delta-like and double-delta kernels has been tested, we proceed to analyze the displacement produced by the kernel $\boldsymbol{k}^{(\delta_{2},r)}$, with $r\neq 0$, where the double-delta function has been considered to avoid any output suppression. The effects of this kernel are tested for $j=20$ and displacements to the right amounting to $12.5\%$ ($r=5$), $25\%$ ($r=10$), and $37.5\%$ ($r=15$) for the inputs $\boldsymbol{u}^{(18)}$ and $\boldsymbol{u}^{(15)}$. For both inputs, displacements of $12.5\%$ and $25\%$ produce an output that displaces without any major distortion, as the right tail of signals vanishes prior to reaching the lattice edge (see Fig.~\ref{fig:filter-1}). This is not the case with displacements of $37.5\%$, where the output becomes perturbed in channels near the edge.  For $j=100$, we consider displacement kernels equivalent to those of the previous case, namely $12.5\%$ ($r=25$), $25\%$ ($r=50$), and $37.5\%$ ($r=75$), as well as inputs with the same number of oscillation, such as $\boldsymbol{u}^{(98)}$ (two nodes) and $\boldsymbol{u}^{(95)}$ (five nodes). Fig.~\ref{fig:filter-2} displays the resulting output, which exhibits only subtle distortions when subjected to $75\%$ displacements. For eigenmodes with the same quantity of oscillations, the tail of both modes is larger than the scenario where $j=20$, which results in reduced edge effects leading to less interference at the edges.

\subsection{Gaussian-like smoothing filter} 
In this section, we consider the $\boldsymbol{u}^{(j)}$ filter and a discrete Gaussian filter as testing kernels to analyze the output for a noisy signal composed by a rectangular window function with added background Gaussian noise. Before proceeding, it is worth comparing the two filters. In order to take them to equal grounds, we generate discrete Gaussian filters with the same standard deviation of the corresponding $\boldsymbol{u}^{(j)}$ filter. For instance, for $j=20$, the standard deviation of $\boldsymbol{u}^{(j)}$ is $\sigma\approx 3.1623$. The corresponding discrete Gaussian filter with the same standard deviation and sampled in the same grid leads to a kernel indistinguishable from the first one. The difference between these two kernels can be quantified through the distance~\eqref{distance}. For $j=20$ and $j=100$, the distances are approximately $0.0111017$ and $0.0010284$, respectively, and become smaller for larger waveguide arrays. This reinforces the use of the filter $\boldsymbol{u}^{(j)}$ as an alternative candidate for a Gaussian filter, which in turn can be generated from another Jx lattice with length $\alpha=\pi/2$ by exciting its $j$ input channel. Nevertheless, contrary to a Gaussian filter, the standard deviation of the filter $\boldsymbol{u}^{(j)}$ is fixed for each $j$ and cannot be modified. This means that narrow or peaked signals might be suppressed during the process, and thus this filter is better suited for smoothing wide-enough signals.

\begin{figure}
\centering
\includegraphics[width=0.48\textwidth]{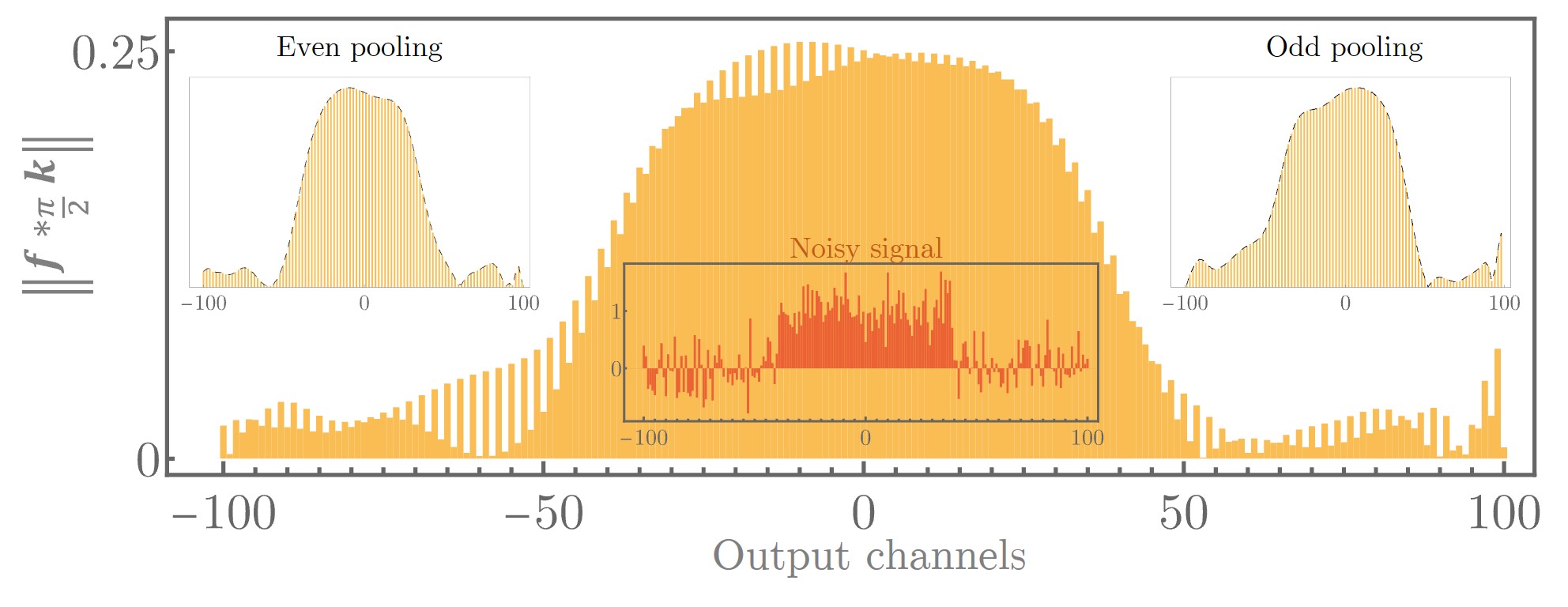}
\caption{Convolution norm $\Vert(\boldsymbol{f}\underset{\pi/2}{\ast}\boldsymbol{k})_{q}\Vert$ using the $\boldsymbol{u}^{(j)}$ filter and a noisy signal $\boldsymbol{f}$ (lower inset) composed of a rectangular window plus Gaussian noise for $j=100$. The upper-left and upper-right insets depict the even and odd pooling, respectively, associated with the convolution output.}
\label{fig:filter-gauss}
\end{figure}

Be the input signal $\boldsymbol{f}$ composed of a rectangular window with added Gaussian noise and the Gaussian-like filter $\boldsymbol{u}^{(j)}$ the inputs for our convolution device. We focus on the case $j=100$ so that each DFrFT element in the device approximates the conventional DFT, and thus results close to the conventional convolution are expected. The output of the latter is depicted in Fig.~\ref{fig:filter-gauss}, where the noisy input signal is also shown (lower inset) for reference. This reflects the fact that this setup works as a smoothing process indeed, which also averages the initial noise but remarks the hidden rectangular window inside the noise. Still, the output contains noise that is inherently introduced from the Jx photonic lattice. As discussed above, any signal and convolution kernel leads to an output that splits into the even and odd channels, depending on the parity of $j$. This classification at the output reveals that an additive term exists in the even output that is not present on the odd one, or vice versa, depending on the parity of $j$ and the input signal $\boldsymbol{f}$ (see Eqs.~(A2)--(A3) of Appendix~\ref{sec:APPA}). 

In filtering processes, it is common to introduce the notion of \textit{pooling layers}, in which the effects of some filtering procedures are sharped by splitting the sampling space into sub-groups and performing specific operations on the latter. As a result, the sampling space of the output of the pooling layer is smaller than the original output. This is particularly common in \textit{convolutional neural networks}, where the pooling process depends on the intended task in hands, such as minimum, maximum, and average pooling\footnote{For maximum pooling, the $N=2m$-dimensional output is decomposed into $m$ sub-groups, each of dimension two, where only the maximum number within each sub-group is preserved.}. Here, we introduce the \textit{even and odd pooling}, an operation that consists of splitting the output signal into its even ($2q$) and odd channels ($2q+1$). That is, for even pooling, only the information of the output channels $2q$ is analyzed, and likewise for the odd pooling. Although the resulting sampling space is half of the original output, the previously described noisy behavior is suppressed. This is particularly illustrated in the upper insets in Fig.~\ref{fig:filter-gauss}. 

\subsection{Edge detection filter} 
Although a large variety of edge detection filters exist, the general idea behind them lies in the smoothing of the signal and the gradient of the same. The smoothing is necessary to avoid unnecessary abrupt changes or singularities of step-like signals. For each smoothing filter, one may associate an edge detection scheme, which might or not be suitable for the signal in question. For the conventional convolution, it is quite a straightforward fact that the differentiation operator commutes with the convolution operation. This means that, instead of determining the gradient of the filtered signal, one can, in principle, compute the convolution using the gradient of the filter kernel instead. This simplifies the complexity of the edge detection scheme. The Gaussian derivative, the first Hermite-Gaussian mode, becomes the most natural edge detection filter as it comprises the Gaussian filtering and the gradient operation required for edge detection. Other alternatives exist in the literature, such as the \textit{Canny}~\cite{Can86} and Laplacian~\cite{Laplace} filters. 

\begin{figure}
\centering
\includegraphics[width=0.48\textwidth]{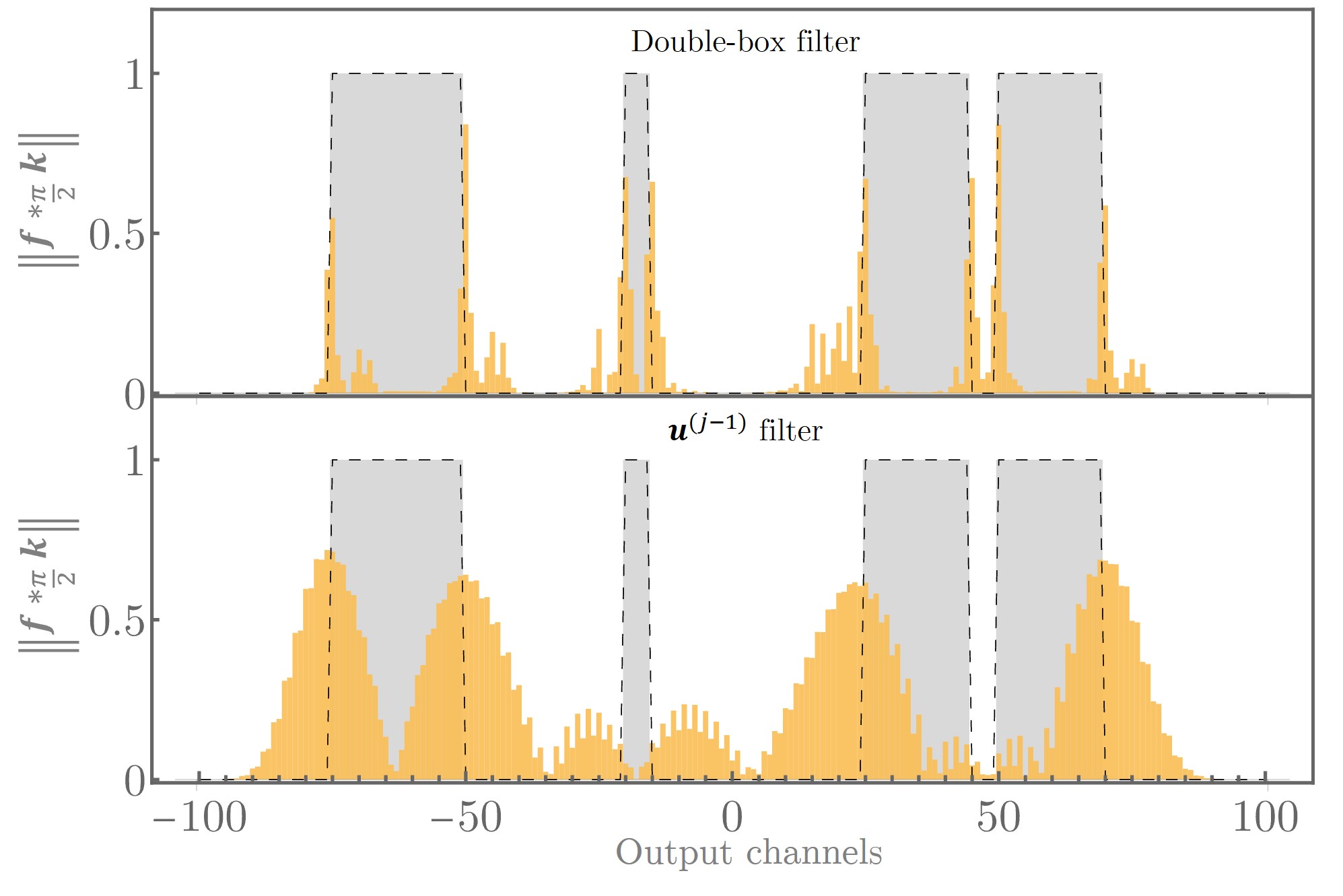}
\caption{Input signal (black-shaded area) and the corresponding convolution norm using the filter $\boldsymbol{u}^{(j-1)}$ (lower row) and double-box (Canny edge) filter $\boldsymbol{k}^{(c)}$ (upper row) for a waveguide array with $j=100$.}
\label{fig:filter-edge-1}
\end{figure}

In our scheme, the eigenmode $\boldsymbol{u}^{(j-1)}$ becomes the natural candidate for the edge detection kernel as it converges to first-order Hermite-Gauss mode in the continuous limit. This can be easily generated by exciting the $(j-1)$-th input of a Jx photonic lattice of length $\pi/2$. One of the drawbacks of such a filter is the lack of control over its width, which limits edge detection in narrow signals. We thus consider the double-box (Canny edge)~\cite{Can86} kernel $\boldsymbol{k}^{(C)}=-\delta_{q,-2}-\delta_{q,-1}+\delta_{q,2}+\delta_{q,3}\equiv (0,\ldots,-1,-1,0 ,0,1,1,\ldots,0)$ as the second filter to perform edge detection. Here, we placed two consecutive zeros at $q=0$ and $q=1$ so that the kernel contains an even number of elements, which, as previously discussed, prevents the output from showing additional noisy behavior for even $j$. 

For the input signal, we devise an ideal sequence of window functions of different widths and spacing. Such an input serves as an enlightening test scenario to understand any potential issues and advantages for each filter, the outcome of which is depicted in Fig.~\ref{fig:filter-edge-1}. In the latter, one can identify a relatively large window at the left of the input. The $\boldsymbol{u}^{(j-1)}$ filter is capable of identifying the edges of the signal in a clean way. Although the double-box filter accomplishes the same task, the signal becomes noisier than the first filter. The situation changes in the intermediate region of the signal, where a narrow window appears. As previously argued, the first filter fails to detect the edges, whereas the double-box filter does the job. Lastly, on the right side of the signal, the first filter spikes up at the outer edges of the signal and fails to detect the inner sides. 

\begin{figure}[h]
\centering
\includegraphics[width=0.48\textwidth]{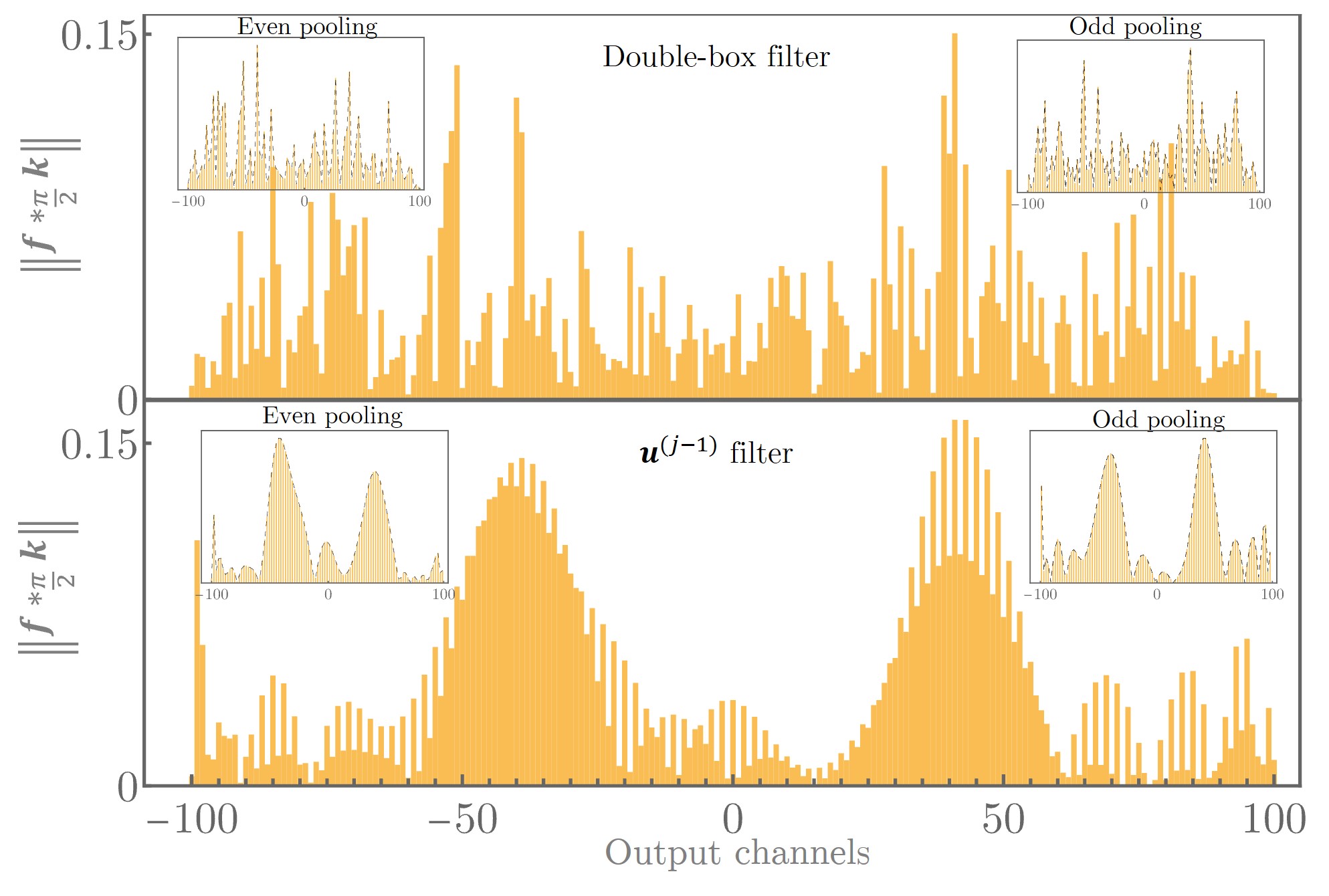}
\caption{Edge detection scheme using the noisy rectangular signal of Fig.~\ref{fig:filter-gauss} and the convolution kernel $\boldsymbol{u}^{(j-1)}$ (lower row) and the double-box filter $\boldsymbol{k}^{(c)}$ (upper row) with $j=100$. In both cases, the upper-left and upper-right insets depict the even and odd pooling of the convolution output, respectively.}
\label{fig:filter-edge-2}
\end{figure}

Thus, the double-box filter provides a more accurate detection scheme but induces extra noise into the output, which is a less-than-ideal scenario when dealing with more complex input signals, such as those with background noise. To verify this, let us consider a simple window signal exposed to a background Gaussian noise with a standard deviation $\sigma=0.3$. We now use this noisy signal to test the edge detection capacities using the previously introduced two filters. The results are shown in Fig.~\ref{fig:filter-edge-2}, where it is clear that the $\boldsymbol{u}^{(j-1)}$ does a better job suppressing the initial background noise and finding the edges of the clean signal. The even and odd pooling at the output improves the signal quality by removing the oscillatory behavior between consecutive even and odd ports, as seen in the upper-left and upper-right insets of Fig.~\ref{fig:filter-edge-2}. The double-box filter produces an output indistinguishable from noise (see the upper-center inset of Fig.~\ref{fig:filter-edge-2}), an issue not fixed with the above-mentioned pooling scheme. As a possible workaround to this problem, one can consider a wider double-box kernel, namely, $\boldsymbol{k}^{(d-C)}:=(0,\ldots,1,1,1,1,0,0,1,1,1,1,\ldots,0)^{T}$. The width of such a modified double-box filter can be steered in order to control the effectiveness of detecting peaked and noisy signals.

\section{Simulations}
For completeness, a wave simulation of the proposed architecture is performed further to test the reliability of the fractional convolution device. To this end, the beam-envelop formulation is used in \texttt{COMSOL Multiphysics} to speed up the design and computational time of the architecture, valid for electromagnetic waves whose oscillations in the propagation direction are much faster than those in the perpendicular direction. In this regime, electric fields take the form $\vec{\mathcal{E}}(\vec{r};t)=\vec{E}(\vec{r}_{T})e^{-i(w t + \vec{k}\cdot\vec{r}_{L})}$, with $\vec{r}_{T}$ and $\vec{r}_{L}$ the position vector on the transverse and longitudinal direction of propagation, respectively. Furthermore, $\vec{k}= \beta \hat{r}_{L}$ is the propagation vector with $\beta$ the corresponding propagation constant. Here, we consider a two-dimensional setup in which the direction of propagation is fixed at $\hat{x}$ so that the transverse electric field takes the form $E(\vec{r})=E(y)\hat{z}$. 

\begin{figure*}
    \centering
    \includegraphics[width=0.85\textwidth]{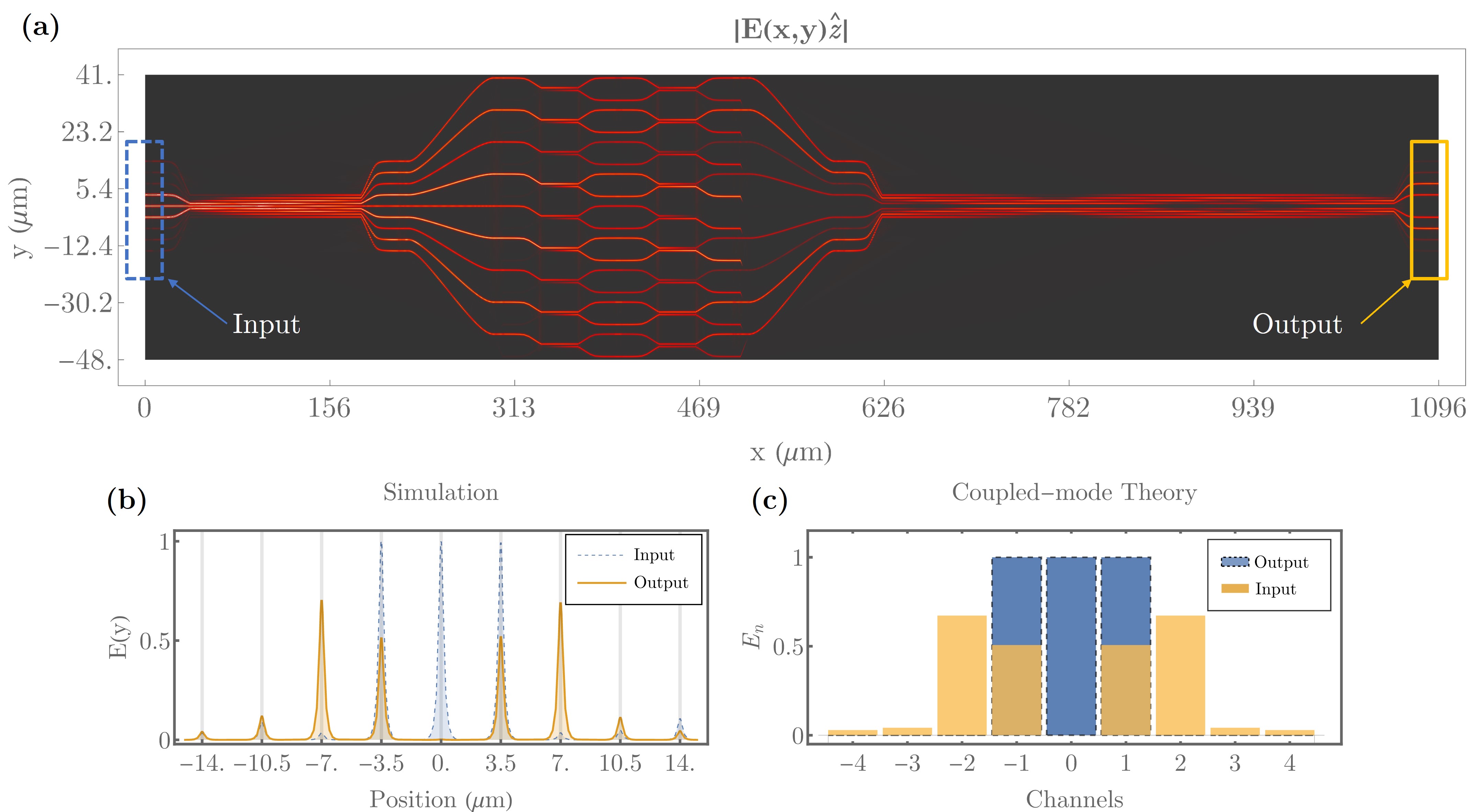}
    \caption{(a) Propagation of the electric field modulus $\vert E(x,y)\hat{z}\vert$ throughout the convolution architecture. (b) Input and output field amplitudes of the simulated results, where the gray-shaded area denotes the location of the waveguide core. (c) Theoretical predictions from coupled mode theory.}
    \label{fig:sim}
\end{figure*}

In the current setup, we consider a 1550 nm infrared source propagating through a silicon waveguide (core). The core has a transverse square shape with $200 nm$ of width and a refractive index of $n_{core}=3.48$, which renders an effective index of $2.85$ for the effective two-dimensional model. The core is surrounded by a fused silica cladding (SiO2) with a refractive index of $n_{cla}=1.47$. This leads to a unique guided mode with propagation constant $\beta=8.9151\times 10^6$ rad/m. The guided mode has a maximum point at the core center that varies slowly within the core and decays exponentially throughout the cladding. To illustrate the complete convolution architecture, we consider a device with $N=9$ channels ($j=4$), where the waveguides corresponding to the channels $p=-1,0,+1$ are uniformly separated by 0.63 nm, whereas the subsequent waveguides are separated by 637.99 nm ($p=1,2$ and $p=-1,-2$), 657.13 nm ($p=2,3$ and $p=-2,-3$), and 700.11 nm ($p=3,4$ and $p=-3,-4$). This separation ensures the coupling strengths required for the $J_x$ lattice so that $\widetilde{\kappa}L=\pi/2$ is fulfilled for the lattice length L=139.32 $\mu m$, which is the required length to perform the DFrFT, and L=419.97 $\mu m$ for the inverse DFrFT. The coupling length and separation of each 50:50 directional coupler are 29.35 $\mu m$ and 630 nm, respectively. Passive phase shifters, such as small bends, are placed before any bend to compensate for any deviation in the optical path, whereas active phase shifters are placed after every 50:50 directional coupler to modulate phase and amplitude throughout the MZI.

Although phase shifters can be deployed through conventional thermo-optical effects~\cite{liu2022thermo,harris2014efficient}, phase-change materials have opened the possibility for even more compact phase shifters. Particularly, an implementation based on Sb2Se3 has shown to achieve a phase shift of approximately $(\pi/11)$ rads/$\mu m$ for a 1550 nm source~\cite{rios2022ultra}. The ongoing simulations use the latter solution for the active phase elements to steer the properties of the MZI. From the previous considerations, and after including proper bends to connect the $J_x$ lattices with the MZI elements and the phase shifters, the total dimension of the optical device becomes 1096 $\mu m \times$ 89 $\mu m$. Clearly, this does not include the packaging, but it provides a reasonable estimation of the device size. As a particular example, the edge detection scheme is considered in our simulations, where the window signal $\boldsymbol{f}=\{0,0,0,1,1,1,0,0,0\}$ is injected at the input and the kernel function $\boldsymbol{k}^{(C)}=\{0,0,-1,-1,0,1,1,0,0\}$ is used as the convolution filter. Recall that the phase shifters of the MZI array are tuned so that they render the DFrFT of $\boldsymbol{k}^{(C)}$, which in this case becomes $\boldsymbol{K}^{(C)}=\{0.935 e^{-i \frac{\pi}{2}} \mathbin{,} 0.935 e^{-i \frac{\pi}{2}} \mathbin{,} 0.353 e^{-i \frac{\pi}{2}} \mathbin{,} 0.353 e^{-i \frac{\pi}{2}} \mathbin{,} 0 \mathbin{,} 0.353 e^{i \frac{\pi}{2}} \mathbin{,} 0.353 e^{i \frac{\pi}{2}} \mathbin{,} 0.935 e^{i \frac{\pi}{2}} \mathbin{,} 0.935 e^{i \frac{\pi}{2}}\}$. Because of the low number of channels, the double-box filter has been used instead of the $\boldsymbol{u}^{(j-1)}$ filter. As discussed earlier, the latter filter is unsuitable for narrow input signals, such as the one considered here. The simulated propagation of the electric field modulus throughout the architecture is depicted in Fig.~\ref{fig:sim}(a). The respective input and output electric field modulus in Fig.~\ref{fig:sim}(b), where the gray shaded area denotes the location of the waveguide core. Clearly, the simulation output reveals the two expected peaks, one next to each edge of the input signal, thus performing the desired edge detection. Furthermore, Fig.~\ref{fig:sim}(c) portrays the theoretical results obtained from coupled mode theory, from which one can notice a good agreement between the simulated and theoretical results. 


\section{Lattice Defects}
Hitherto, our architecture has been tested for different filtering processes and proved reliable even for noisy signals. Still, errors may appear during the manufacturing process of the Jx lattice, rendering waveguide arrays that deviate from the ideal one described by the Hamiltonian $\mathbb{H}$. Given the nature of the unpredicted errors, we can introduce a new lattice model accounting for such effects through added random noise. To this end, let us consider the new perturbed lattice Hamiltonian $\widetilde{\mathbb{H}}(\delta)=\mathbb{H}+\delta \mathbb{R}$, with $\delta>0$ a strength parameter and $\mathbb{R}$ a $(2j+1)\times(2j+1)$ a symmetric matrix whose components $\mathbb{R}_{p,q}$ are randomly distributed numbers in the interval $(-1,1)$. In this form, the wave propagation is still ruled by a lossless process characterized by the unitary evolution operator $\widetilde{\mathbb{G}}(\alpha;\delta)=e^{-i\alpha\widetilde{\mathbb{H}}(\delta)}$. Since the DFrFT of the convolutional kernel $\boldsymbol{k}$ is programmed through the MZI array, we only require the implementation of two Jx lattices (see Fig.~\ref{fig:device}), and thus defects are only considered in such layers. For each one of the latter, we consider a different perturbation matrix $\mathbb{R}$ so that defects may not be the same. Recall that the second DFrFT layer has a different length ($2\pi-\alpha$), corresponding to the inverse DFrFT operation. Following the same treatment as in previous sections, we fix $\alpha=\pi/2$ throughout the rest of this section. 

\begin{figure}[h]
\centering
\includegraphics[width=0.48\textwidth]{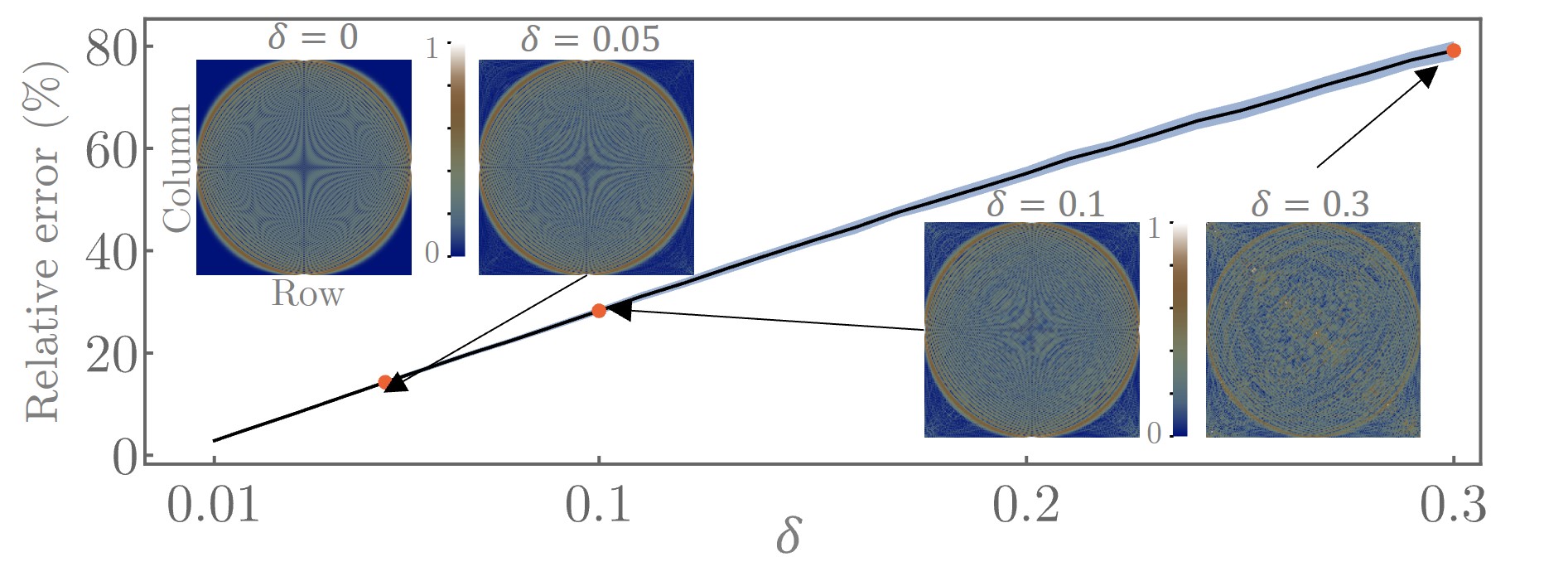}
\caption{Mean (line) and standard deviation (shaded area) of the relative percent error ($E(\delta)\times 100\%$) for a set of 100 randomly generated perturbed DFrFT matrices ($\mathbb{G}(\pi/2;\delta)$) per value of $\delta$. The insets depict $\vert \mathbb{G}(\pi;\delta)\vert$ (matrix modulus) of a random sample out of the set of 100 random matrices for $\delta=0.05,0.1,0.3$.}
\label{fig:error-1}
\end{figure}

In order to understand the influence of the perturbation $\mathbb{R}$, we compute the relative error $E(\delta):=\Vert \widetilde{\mathbb{G}}(\pi/2;\delta)-\mathbb{G}(\pi/2)\Vert_{F}/\Vert \mathbb{G}(\pi/2) \Vert_{F}$, where $\Vert\mathbb{M}\Vert_{F}:=\sqrt{\operatorname{tr}(\mathbb{M}\mathbb{M}^{\dagger})}$ stands for the Frobenius norm of the complex matrix $\mathbb{M}$. The latter allows quantifying any deviation from the original DFrFT as a function of the perturbation strength $\delta$. To avoid any misleading behavior, we compute 100 perturbed DFrFT matrices ($\mathbb{G}(\pi/2;\delta)$) and their corresponding relative percent error ($E(\delta)\times 100\%$) for each $\delta$, we then compute the mean and standard deviation per $\delta$. The results are shown in Fig.~\ref{fig:error-1}, where it can be seen that a relative error of $14.7\%$ is reached for $\delta=0.05$, whereas the error rises to $28\%$ and $80\%$ for $\delta=0.1$ and $\delta=0.3$, respectively. A graphic visualization of the DFrFT and its perturbed counterparts is presented as the matrix modulus ($\vert \mathbb{G}(\pi/2;\delta)\vert_{p,q}$=$\vert \mathbb{G}_{p,q}(\pi/2;\delta)\vert$) in the insets of Fig.~\ref{fig:error-1}. These figures reveal a mild distortion of the DFrFT for $\delta=0.05$ and $\delta=0.1$, and an indistinct form for $\delta=0.3$. We thus should expect similar results for relative errors up to approximately $28\%$ ($\delta\approx 0.1$).

\begin{figure}[h]
\centering
\subfigure[$\boldsymbol{u}^{(j-1)}$ filter]{\includegraphics[width=0.48\textwidth]{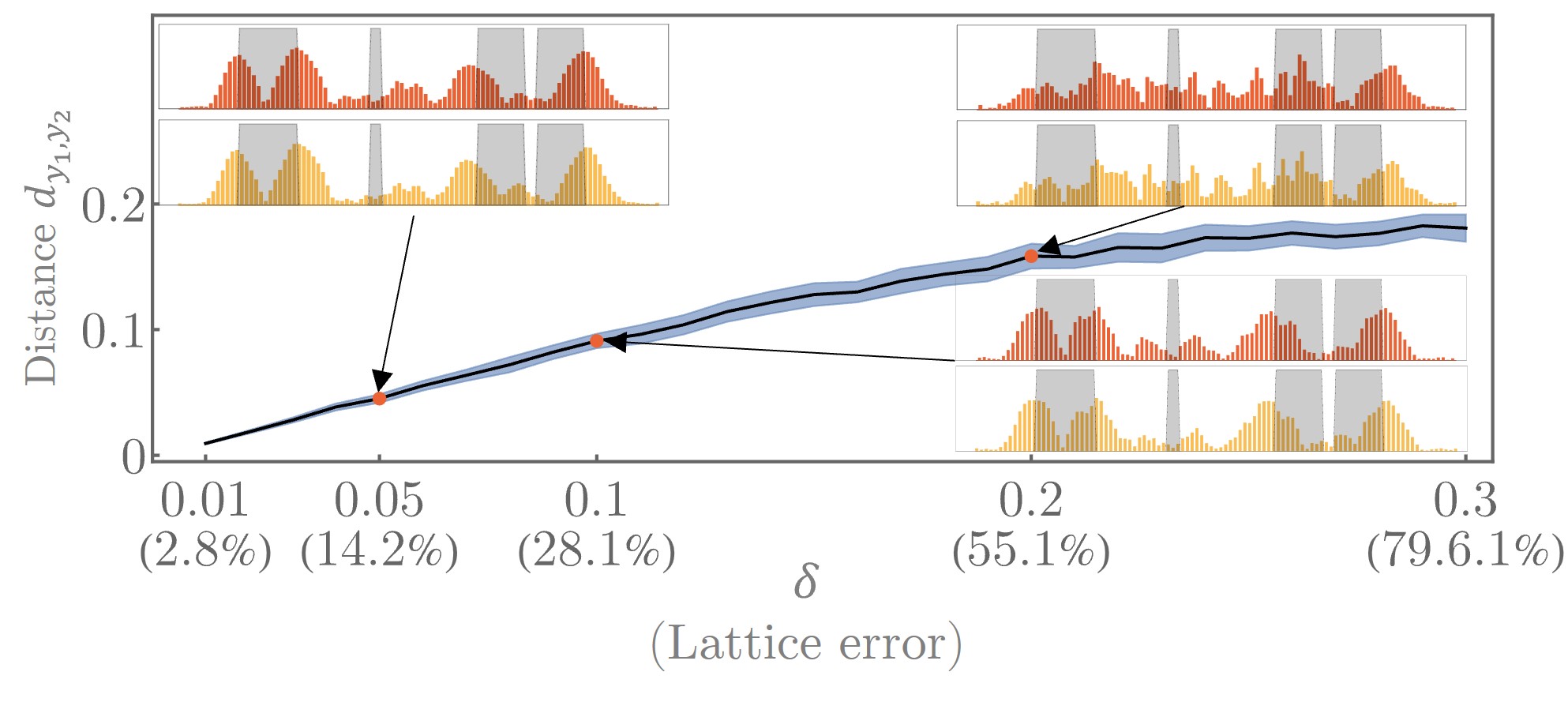}
\label{fig:error-edge}}
\\
\subfigure[Double-box filter]{\includegraphics[width=0.48\textwidth]{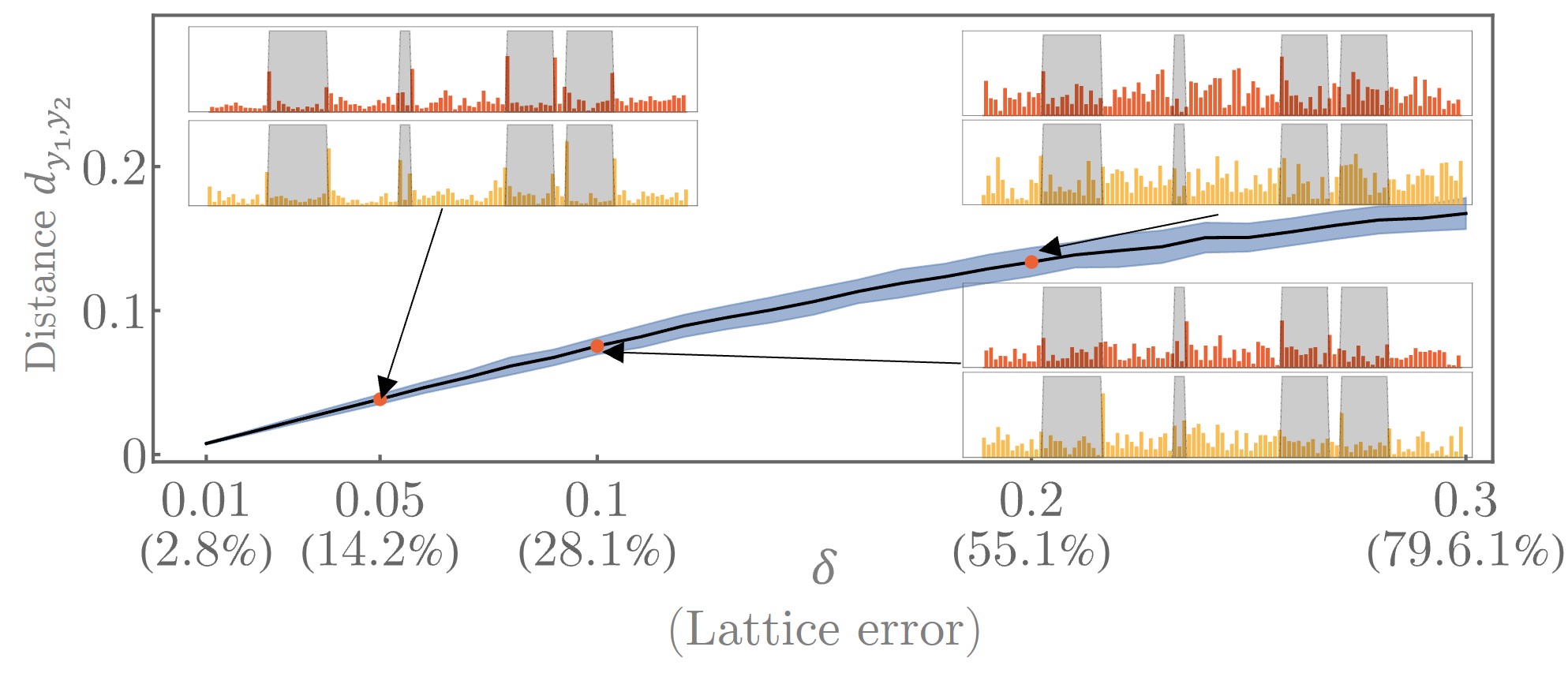}
\label{fig:error-canny}}
\caption{Distance function ($d_{y_1,y_2}(\delta)$) between the unperturbed ($\delta=0$) and perturbed ($\delta\neq 0$) convolution operations using the $\boldsymbol{u}^{(j-1)}$~(a) and double-box (b) filters. The plot is presented as a function of $\delta$ and the lattice error $E(\delta)\times 100\%$. The mean (line) and standard deviation (shaded area) are computed by generating a set of 100 random perturbed DFrFT matrices ($\mathbb{G}(\pi/2;\delta)$) per value of $\delta$. The insets depict the output of the perturbed filtering using the even (yellow) and odd (red) pooling operation for $\delta=0.05,0.1,0.2$.}
\label{fig:error-2}
\end{figure}

Notably, we can determine the effects of the lattice defects on the edge detection scheme using both the $\boldsymbol{u}^{(j-1)}$ and double-box filters. Since the output of the perturbed convolution is also a vector in $\mathbb{C}^{2j+1}$ and we are interested in deviations on the output vector compared to the ideal case, we use the distance function~\eqref{distance} with $y_{1}$ and $y_{2}$ the perturbed an unperturbed outputs. A set of 100 random matrices $\mathbb{G}(\pi/2;d)$ and another one for $\mathbb{G}(3\pi/2;d)$ are generated so that multiple edge detection operations can be computed for each $\delta$. The corresponding distance is computed for every output with respect to the unperturbed one; the mean and standard deviations for each $\delta$ are shown in Fig~\ref{fig:error-2}. The distance using both filters leads to similar results, where it can be noticed that lattice errors of $14.2\%$ do not produce any relevant change in the edge detection for both filters (see insets in Figs.~\ref{fig:error-edge}-\ref{fig:error-canny}). For lattice errors around $28.1\%$, the filter $\boldsymbol{u}^{(j-1)}$ is still performing well, whereas the double-box filter misses some of the edges during the detection scheme. In the latter, the even and odd pooling allows for a cleaner output where the edges are properly identified. This highlights the relevance of such pooling operations in this convolution architecture in the presence of lattice defects. In turn, the output is indistinct for an error of $55.1\%$, where the pooling operations also fail to extract any relevant information. 

\section{Conclusions} 
The DFrFT based on the Jx photonic lattice proved to be a valuable resource for constructing an adequate convolution architecture. This choice was not fortuitous as the asymptotic behavior for large channel numbers reveals that such a lattice asymptotically reproduced the continuous FT operation. Interestingly, the same lattice can be used as a source to generate convolution kernels that can be later used in the convolution process. So far, the $j$-th and $(j-1)$-th modes have proved to be suitable kernels to produce Gaussian-like smoothing and edge detection tasks, respectively. As a test setup, we used a single delta kernel in the middle channel of the convolution kernel, which, in the conventional convolution, leaves the input invariant. To quantify any difference, the distance was computed between the lattice eigenmodes at the input and their respective convolution output, which showed that eigenmodes around $n=0$ are more prompt to deviate at the output. In turn, eigenmodes close to the edges are less affected, and their deviation decreases when a larger number of channels is considered.

For more general signals without any particular symmetry, the output decomposes into an even and odd channel output $(\boldsymbol{f}\underset{\pi/2}{\ast}\boldsymbol{k}^{(\delta)})_{2q}$ and $(\boldsymbol{f}\underset{\pi/2}{\ast}\boldsymbol{k}^{(\delta)})_{2q+1}$, respectively. The latter can be written entirely in terms of $\mathcal{C}_{m}$ and the basis elements $\boldsymbol{u}^{(m)}$ (see Appendix~\ref{sec:APPA}), from which one notices that even and odd channels at the output differ by an additive term that breaks any potential continuity contiguous channels. This explains the wavy behavior of the convolution when noisy signals are considered, such as those depicted in Fig.~\ref{fig:filter-gauss} and Fig.~\ref{fig:filter-edge-2}, and also justified the introduction of even and odd pooling as a mechanism to clean the convolution output.

Interestingly, the architecture has shown to be a reliable source for performing edge detection tasks for both clean and noisy inputs. This is particularly achieved by either exploiting one of the lattice eigenmodes $\boldsymbol{u}^{(j-1)}$ as an edge-detection filter or using the conventional double-box filter. The former one suits better for signals whose width is at least of the order of the eigenmode $\boldsymbol{u}^{(j)}$ width, whereas the double-box filter works better for narrow signals. This fact is further tested by computing wave simulations of the full architecture, which reveals a good match with the theoretical predictions. The simulation design reveals some challenges in achieving the ideal Jx lattice, mainly as undesired coupling of the waveguides during the bends connecting the waveguide array to the other elements of the device. Numerical data from the simulation suggest that such undesired coupling occurs around $1 \mu m -2 \mu m$ prior to the array end points for an initial waveguide separation of $630 nm$ in the central channels. This separation has been chosen as the total propagation length to achieve the DFrFT is $139.32 \mu m$. This is much larger than the distance where the undesired coupling occurs, and any potential deviations from the ideal behavior do not compromise the desired functionality.

Our convolutional architecture has shown resilience due to defects on the Jx lattice. This was elucidated by adding a random and symmetric interaction to the lattice Hamiltonian so that the wave evolution through the lattice is still unitary despite the included random defects. The intensity of such defects is controlled by employing a strength parameter $\delta$. The error induced in the wave evolution is relatively large even for $\delta\approx 0.1$. This was further illustrated by performing the edge detection operation, which is still reliable with lattice errors of up to $28\%$. Thus, the coupling parameters $\kappa_{p}$ are allowed to have mild deviations from their exact values and the architecture is still expected to perform the convolution task.

For completeness, it is worth mentioning that one can alternatively construct a convolution scheme based on circulant convolutions. The convolution kernel has been replaced by a circulant matrix that multiplies the input signal, where the circulant matrix factorizes in terms of the DFT. Such an approach can be easily modified and implemented using the evolution operator~(\ref{G-alpha}), as proved in Appendix~\ref{sec:APPB}. Nevertheless, the implementation of the latter in the form of a photonic architecture is unclear, and the approach discussed throughout this work seems more adequate from a practical point of view, as it is possible to identify each of the matrices with the optical passive elements.

\begin{acknowledgments}
This project is supported by the U.S. Air Force Office of Scientific Research (AFOSR) Young Investigator Program (YIP) Award\# FA9550-22-1-0189 and by CUNY's Junior Faculty Research Award in Science and Engineering funded by the Alfred P. Sloan Foundation.
\end{acknowledgments}

\section*{Disclosures}
Patent pending.


\appendix

\section{Convolution expansion}
\label{sec:APPA}

\begin{ruledtabular}
\begin{table*}
\centering
\begin{tabular}{c|c|c|c|c|c|c|c}
$q$ & $\ell$ & s & $(\boldsymbol{u}^{(\ell)}\underset{\pi/2}{\ast}\boldsymbol{e}^{(s)})_{q}$ & $q$ & $\ell$ & s & $(\boldsymbol{u}^{(\ell)}\underset{\pi/2}{\ast}\boldsymbol{e}^{(s)})_{q}$ \\
\hline
\multirow{4}{*}{$2q$} & \multirow{2}{*}{$2\ell$} & \small{$2s$} & \small{$\mu_{2\ell,2s}^{(2q)}+\sigma_{2\ell,2s}^{(2q)}$} & \multirow{4}{*}{$2q+1$} & \multirow{2}{*}{$2\ell$} & \small{$2s$} & \small{$0$} \\
 & & \small{$2s+1$} & \small{$0$} &  & & \small{$2s+1$} & \small{$\sigma_{2\ell,2s+1}^{(2q+1)}$} \\
 & \multirow{2}{*}{$2\ell+1$} & $2s$ & $0$ &  & \multirow{2}{*}{$2\ell+1$} & $2s$ & $\sigma_{2\ell+1,2s}^{(2q+1)}$ \\
 & & $2s+1$ & $\sigma_{2\ell+1,2s+1}^{(2q)}$ &  & & $2s+1$ & $0$
\end{tabular}
\caption{Convolution at the output channel $q$ with the input $\boldsymbol{u}^{(\ell)}$ and the convolution kernel $\boldsymbol{e}^{(s)}$.}
\label{tab:tab1}
\end{table*}
\end{ruledtabular}

In this section, we explicitly compute the output at the convolution device for any input signal and convolution kernel to explain the oscillatory-like behavior produced by the device. This is clearly seen if one rewrites the input signal using the photonic Jx lattice eigenmodes and the kernel in terms of delta-kicks. This is indeed a legitimate operation as we can map between any basis through unitary operations, and the result is independent of the basis. We consider the most general form $\boldsymbol{f}=\sum_{\ell} f_{\ell}\boldsymbol{u}^{(\ell)}$ for the input signal and $\boldsymbol{k}=\sum_{s}k_{s}\boldsymbol{e}^{(s)}$ for the kernel, with $\boldsymbol{e}^{(s)}$ the $s$-th canonical unit vector in $\mathbb{C}^{N}$. Although the latter expansion of the signal and kernel is not unique, it is useful and straightforward to explain the behavior of the fractional convolution.

Since our convolution scheme defines a linear operation, we can start by computing the output for $\boldsymbol{f}=\boldsymbol{u}^{(\ell)}$ and $\boldsymbol{k}=\boldsymbol{e}^{(s)}$, whose Fourier transform leads to $\boldsymbol{F}^{(\ell)}_{p}=(-i)^{\ell}\boldsymbol{u}^{(\ell)}_{p}$ and $\boldsymbol{K}^{(s)}_{p}=(-i)^{s-p}\boldsymbol{u}^{(s)}_{p}$, respectively. The convolution is then determined from Eq.~\eqref{Conv-1} by computing the inverse Fourier transform $\boldsymbol{F}^{(\ell)}_{p}\boldsymbol{K}_{p}^{(s)}$. This is simplified by recalling the identity (see Methods section of~\cite{Wei16}) $G_{pq}(\pi/2)=(-i)^{q-p}u^{(q)}_{p}$, which in our case can be cast into the form $G_{qp}(-\pi/2)=(-i)^{p-q}u^{(q)}_{p}$, where we have used the fact that $u^{(r)}_{s}=(-1)^{s-r}u^{(s)}_{r}$. The straightforward calculations show that the solution can be decomposed as 
\begin{equation*}
\left( \boldsymbol{u}^{(\ell)}\underset{\pi/2}{\ast}\boldsymbol{e}^{(s)} \right)_{q}=\mu_{\ell,s}^{(q)}+\sigma_{\ell,s}^{(q)}
\end{equation*}
where
\begin{subequations}
\begin{eqnarray}
\label{mu}
\mu_{\ell,s}^{(q)}&=&(-i)^{q+\ell+s}u_{0}^{(q)}u_{0}^{(\ell)}u_{0}^{(s)} , \\
\label{sigma}
\sigma_{\ell,s}^{(q)}&=&(-i)^{q+\ell+s} \\
& &\times\sum_{p=1}^{j}\left( u_{p}^{(q)}u_{p}^{(\ell)}u_{p}^{(s)}+u_{-p}^{(q)}u_{-p}^{(\ell)}u_{-p}^{(s)} \right) .
\end{eqnarray}
\end{subequations}

Since the photonic lattice eigenmodes are parity symmetric, $u^{(n)}_{-q}=(-1)^{n}u^{(n)}_{q}$ for even $j$, the function $\mu_{\ell,s}^{(q)}$ survives only for even $q$, $\ell$ and $s$, and vanishes otherwise. Likewise, $\sigma_{\ell,s}^{(q)}$ cancels out according to the parity of the indexes $q$, $\ell$, and $s$, as summarized in Table~\ref{tab:tab1}. Consequently, the convolution Eq.~\eqref{Conv-1} produces an output only on the even or odd channels, as depicted in the cases portrayed in Fig.~\ref{fig:filter-1}.

We can expand the latter results and use them for a general input signal and convolution kernel. From linearity arguments and after some calculation, it is clear that the output at the convolution device takes the form
\begin{multline}
\label{even}
\left(\boldsymbol{f}\underset{\pi/2}{\ast}\boldsymbol{k}\right)_{2q}=\sum_{\ell=-j/2}^{j/2}\sum_{s=-j/2}^{j/2}f_{2\ell}k_{2s}\left(\mu_{2\ell,2s}^{(2q)}+\sigma_{2\ell,2s}^{(2q)} \right)+\\
\sum_{\ell=-(j-1)/2}^{(j-1)/2}\sum_{s=-(j-1)/2}^{(j-1)/2}f_{2\ell+1}k_{2s+1}\sigma_{2\ell+1,2s+1}^{(2q)}
\end{multline}
and
\begin{multline}
\label{odd}
\left(\boldsymbol{f}\underset{\pi/2}{\ast}\boldsymbol{k}\right)_{2q+1}=\sum_{\ell=-j/2}^{j/2}\sum_{s=-(j-1)/2}^{(j-1)/2}f_{2\ell}k_{2s+1}\sigma_{2\ell,2s+1}^{(2q+1)} +\\
\sum_{\ell=-(j-1)/2}^{(j-1)/2}\sum_{s=-j/2}^{j/2}f_{2\ell+1}k_{2s}\sigma_{2\ell+1,2s}^{(2q+1)}
\end{multline}
for even and odd channels, respectively.

From the explicit form of the eigenmodes $u^{(n)}_{q}\equiv (-1)^{q-n}u^{(q)_{n}}$, one can see that Eq.~\eqref{even} and Eq.~\eqref{odd} define continuous functions separately for $q\in\mathbb{R}$. The same behavior is expected for discrete $q$ for each of those functions. When considering the total output, we notice that the even output has an additive term $\mu_{2\ell,2s}^{(2q)}$ (first line in~\eqref{even}) that depends on $q$ and it is not present in the odd output. This term produces a sharp jump when transiting between even and odd channels when the general input signal does not have any parity symmetry, as depicted in the convolution examples presented throughout the main text. 

In turn, by suppressing the total output into its even and odd channels, we eliminate the oscillatory-like behavior and preserve only the relevant components Eqs.~\eqref{even}-\eqref{odd}, which are each smooth in the continuous case. This justifies the smooth behavior at the output when using the even and odd pooling introduced in the main text.

\section{Exact Circulant Convolution Through FrDFT}
\label{sec:APPB}
The convolution presented in the main text is the most immediate scheme that can be implemented by simply using photonic Jx lattices of different lengths. This converges to the continuous convolution in the proper limit $j\rightarrow\infty$. Still, one can rewrite the conventional convolution matrix in terms of the Jx operator by exact means. Although the latter provides no insight into a physical implementation, it is worth discussing the method here.

The convolution operation for two vectors $\boldsymbol{f}$ and $\boldsymbol{k}$ writes as $(\boldsymbol{f}\ast\boldsymbol{k})_{q}=\sum_{n=-j}^{j}\boldsymbol{f}_{q'}\boldsymbol{k}_{q-q'}$, whereas the circulant convolution (defined for cyclic functions $\boldsymbol{k}_{q\pm N}=\boldsymbol{k}_{q}$) can be as $(\boldsymbol{f}\ast\boldsymbol{k})\equiv \mathbb{C}\boldsymbol{f}$, where $\mathbb{C}$ is the circulant matrix 
\begin{equation*}
\boldsymbol{C}=\begin{pmatrix}
\boldsymbol{k}_{1} & \boldsymbol{k}_{N} & \boldsymbol{k}_{N-1} & \cdots & \boldsymbol{k}_{2} \\
\boldsymbol{k}_{2} & \boldsymbol{k}_{1} & \boldsymbol{k}_{N} & \cdots & \boldsymbol{k}_{3} \\
\boldsymbol{k}_{3} & \boldsymbol{k}_{2} & \boldsymbol{k}_{1} & \cdots & \boldsymbol{k}_{4} \\
 & \vdots & & & \vdots \\
\boldsymbol{k}_{N} & \boldsymbol{k}_{N-1} & \boldsymbol{k}_{N-2} & & \boldsymbol{k}_{1}
\end{pmatrix}
.
\end{equation*}
which is clearly circulant.

The latter can be rewritten by using the fact that any circulant matrix can be decomposed as the product of DFT matrices and a diagonal matrix, the components of which are the components of the DFT of the kernel in question. That is, $\boldsymbol{C}=\boldsymbol{F}\boldsymbol{D}\boldsymbol{F}^{\dagger}$, with $\boldsymbol{C}_{p,q}=N^{-1/2}e^{-2\pi i \frac{pq}{N}}$ the DFT matrix and $D=diag(\kappa_{1},\ldots, \kappa_{N})$ with $\mathcal{F}[\boldsymbol{k} ]_{p}=\kappa_{p}$. We can always rewrite the DFT matrix in terms of the $G_{p,q}(\pi/2)$ matrix of the photonic lattice Jx as shown below.

First, let us recall that the DFT matrix has the set of degenerate eigenvalues $\{1,-1,i,-i\}$ for $N>4$, where the exact degeneracy depends on whether $N=4m,4m+1,4m+2,4m+3$ for some positive integer $m$ (see~\cite{Dic82} for an extensive discussion). Despite this degeneracy, it is possible to find an orthonormal set of eigenvectors $\{\boldsymbol{v}^{(n)}\}_{n=1}^{N}$ that are complete in the vector space $\mathbb{C}^{N}$. This set is usually computed by identifying a matrix that commutes with $\boldsymbol{F}$, as discussed in~\cite{Dic82}; see also~\cite{Bos01}. Such a basis can always be arranged in such a way that the spectral decomposition $\boldsymbol{F}=\sum_{n=1}^{N-1}(-i)^{n}\boldsymbol{v}^{(n)}[\boldsymbol{v}^{(n)}]^{\dagger}$ is allowed. Likewise, the set of eigenvectors of the FrDFT, $\{\boldsymbol{u}^{(n)}\}$, also forms an orthonormal in $\mathbb{C}^{N}$. Thus, one can move from one basis to the other through the unitary operation $\boldsymbol{v}^{(n)}=\mathcal{O}\boldsymbol{u}^{(n)}$, where $\mathcal{O}=\sum_{n=1}^{N}\boldsymbol{v}^{(n)}[\boldsymbol{u}^{(n)}]^{\dagger}$. 

In this form, the mapping between the basis combined with the spectral decomposition of the DFT and DFrDFT allows rewriting the circulant convolution matrix in terms of the DFrFT through
\begin{equation}
\label{conv-G}
\boldsymbol{C}=\boldsymbol{F}\boldsymbol{D}\boldsymbol{F}^{\dagger}\equiv \mathcal{O}^{\dagger}\boldsymbol{G}(\tfrac{\pi}{2})\widetilde{\boldsymbol{D}}\boldsymbol{G}(-\tfrac{\pi}{2})\mathcal{O}, \quad \widetilde{\boldsymbol{D}}=\mathcal{O}\boldsymbol{D}\mathcal{O}^{\dagger} .
\end{equation}
The new representation in terms of the DFrFT requires the specific lattice length $\alpha=\pi/2$ to match the eigenvalues of the lattice with those of the DFT. The convolution matrix is now written in terms of the unitary operator $\mathcal{O}$ and the non-diagonal matrix $\widetilde{\boldsymbol{D}}$. The conventional convolution matrix is a particular case of Toeplitz matrices, which are asymptotically approximated for large matrices through circulant matrices. Thus, the circulant convolution and the convolution matrix in Eq.~\eqref{conv-G} approximate the conventional convolution for large $N$.


\bibliography{biblio}

\providecommand{\noopsort}[1]{}\providecommand{\singleletter}[1]{#1}%
\begin{thebibliography}{46}%
\makeatletter
\providecommand \@ifxundefined [1]{%
 \@ifx{#1\undefined}
}%
\providecommand \@ifnum [1]{%
 \ifnum #1\expandafter \@firstoftwo
 \else \expandafter \@secondoftwo
 \fi
}%
\providecommand \@ifx [1]{%
 \ifx #1\expandafter \@firstoftwo
 \else \expandafter \@secondoftwo
 \fi
}%
\providecommand \natexlab [1]{#1}%
\providecommand \enquote  [1]{``#1''}%
\providecommand \bibnamefont  [1]{#1}%
\providecommand \bibfnamefont [1]{#1}%
\providecommand \citenamefont [1]{#1}%
\providecommand \href@noop [0]{\@secondoftwo}%
\providecommand \href [0]{\begingroup \@sanitize@url \@href}%
\providecommand \@href[1]{\@@startlink{#1}\@@href}%
\providecommand \@@href[1]{\endgroup#1\@@endlink}%
\providecommand \@sanitize@url [0]{\catcode `\\12\catcode `\$12\catcode
  `\&12\catcode `\#12\catcode `\^12\catcode `\_12\catcode `\%12\relax}%
\providecommand \@@startlink[1]{}%
\providecommand \@@endlink[0]{}%
\providecommand \url  [0]{\begingroup\@sanitize@url \@url }%
\providecommand \@url [1]{\endgroup\@href {#1}{\urlprefix }}%
\providecommand \urlprefix  [0]{URL }%
\providecommand \Eprint [0]{\href }%
\providecommand \doibase [0]{https://doi.org/}%
\providecommand \selectlanguage [0]{\@gobble}%
\providecommand \bibinfo  [0]{\@secondoftwo}%
\providecommand \bibfield  [0]{\@secondoftwo}%
\providecommand \translation [1]{[#1]}%
\providecommand \BibitemOpen [0]{}%
\providecommand \bibitemStop [0]{}%
\providecommand \bibitemNoStop [0]{.\EOS\space}%
\providecommand \EOS [0]{\spacefactor3000\relax}%
\providecommand \BibitemShut  [1]{\csname bibitem#1\endcsname}%
\let\auto@bib@innerbib\@empty
\bibitem [{\citenamefont {Oppenheim}\ \emph {et~al.}(1997)\citenamefont
  {Oppenheim}, \citenamefont {Willsky}, \citenamefont {Nawab},\ and\
  \citenamefont {Ding}}]{oppenheim1997signals}%
  \BibitemOpen
  \bibfield  {author} {\bibinfo {author} {\bibfnamefont {A.~V.}\ \bibnamefont
  {Oppenheim}}, \bibinfo {author} {\bibfnamefont {A.~S.}\ \bibnamefont
  {Willsky}}, \bibinfo {author} {\bibfnamefont {S.~H.}\ \bibnamefont {Nawab}},\
  and\ \bibinfo {author} {\bibfnamefont {J.-J.}\ \bibnamefont {Ding}},\
  }\href@noop {} {\emph {\bibinfo {title} {Signals and systems}}},\
  Vol.~\bibinfo {volume} {2}\ (\bibinfo  {publisher} {Prentice Hall Upper
  Saddle River, NJ},\ \bibinfo {year} {1997})\BibitemShut {NoStop}%
\bibitem [{\citenamefont {Goodfellow}\ \emph {et~al.}(2016)\citenamefont
  {Goodfellow}, \citenamefont {Bengio},\ and\ \citenamefont
  {Courville}}]{goodfellow2016deep}%
  \BibitemOpen
  \bibfield  {author} {\bibinfo {author} {\bibfnamefont {I.}~\bibnamefont
  {Goodfellow}}, \bibinfo {author} {\bibfnamefont {Y.}~\bibnamefont {Bengio}},\
  and\ \bibinfo {author} {\bibfnamefont {A.}~\bibnamefont {Courville}},\
  }\href@noop {} {\emph {\bibinfo {title} {Deep learning}}}\ (\bibinfo
  {publisher} {MIT press},\ \bibinfo {year} {2016})\BibitemShut {NoStop}%
\bibitem [{\citenamefont {Lugt}(1964)}]{Lugt1964}%
  \BibitemOpen
  \bibfield  {author} {\bibinfo {author} {\bibfnamefont {A.}~\bibnamefont
  {Lugt}},\ }\bibfield  {title} {\bibinfo {title} {Signal detection by complex
  spatial filtering},\ }\href {https://doi.org/10.1109/TIT.1964.1053650}
  {\bibfield  {journal} {\bibinfo  {journal} {IEEE Transactions on Information
  Theory}\ }\textbf {\bibinfo {volume} {10}},\ \bibinfo {pages} {139} (\bibinfo
  {year} {1964})}\BibitemShut {NoStop}%
\bibitem [{\citenamefont {Weaver}\ and\ \citenamefont
  {Goodman}(1966)}]{Weaver:66}%
  \BibitemOpen
  \bibfield  {author} {\bibinfo {author} {\bibfnamefont {C.~S.}\ \bibnamefont
  {Weaver}}\ and\ \bibinfo {author} {\bibfnamefont {J.~W.}\ \bibnamefont
  {Goodman}},\ }\bibfield  {title} {\bibinfo {title} {A technique for optically
  convolving two functions},\ }\href {https://doi.org/10.1364/AO.5.001248}
  {\bibfield  {journal} {\bibinfo  {journal} {Appl. Opt.}\ }\textbf {\bibinfo
  {volume} {5}},\ \bibinfo {pages} {1248} (\bibinfo {year} {1966})}\BibitemShut
  {NoStop}%
\bibitem [{\citenamefont {Davis}\ \emph {et~al.}(1989)\citenamefont {Davis},
  \citenamefont {Waring}, \citenamefont {Bach}, \citenamefont {Lilly},\ and\
  \citenamefont {Cottrell}}]{davis1989compact}%
  \BibitemOpen
  \bibfield  {author} {\bibinfo {author} {\bibfnamefont {J.~A.}\ \bibnamefont
  {Davis}}, \bibinfo {author} {\bibfnamefont {M.~A.}\ \bibnamefont {Waring}},
  \bibinfo {author} {\bibfnamefont {G.~W.}\ \bibnamefont {Bach}}, \bibinfo
  {author} {\bibfnamefont {R.~A.}\ \bibnamefont {Lilly}},\ and\ \bibinfo
  {author} {\bibfnamefont {D.~M.}\ \bibnamefont {Cottrell}},\ }\bibfield
  {title} {\bibinfo {title} {Compact optical correlator design},\ }\href
  {https://doi.org/10.1364/AO.28.000010} {\bibfield  {journal} {\bibinfo
  {journal} {Applied optics}\ }\textbf {\bibinfo {volume} {28}},\ \bibinfo
  {pages} {10} (\bibinfo {year} {1989})}\BibitemShut {NoStop}%
\bibitem [{\citenamefont {Bauchert}\ and\ \citenamefont
  {Serati}(1998)}]{bauchert1998data}%
  \BibitemOpen
  \bibfield  {author} {\bibinfo {author} {\bibfnamefont {K.~A.}\ \bibnamefont
  {Bauchert}}\ and\ \bibinfo {author} {\bibfnamefont {S.~A.}\ \bibnamefont
  {Serati}},\ }\bibfield  {title} {\bibinfo {title} {Data-flow architecture for
  high-speed optical processors},\ }in\ \href
  {https://doi.org/https://doi.org/10.1117/12.304789} {\emph {\bibinfo
  {booktitle} {Optical Pattern Recognition IX}}},\ Vol.\ \bibinfo {volume}
  {3386}\ (\bibinfo {organization} {SPIE},\ \bibinfo {year} {1998})\ pp.\
  \bibinfo {pages} {50--58}\BibitemShut {NoStop}%
\bibitem [{\citenamefont {Lu}\ \emph {et~al.}(2005)\citenamefont {Lu},
  \citenamefont {Hughlett}, \citenamefont {Zhou}, \citenamefont {Chao},\ and\
  \citenamefont {Hanan}}]{lu2005neural}%
  \BibitemOpen
  \bibfield  {author} {\bibinfo {author} {\bibfnamefont {T.~T.}\ \bibnamefont
  {Lu}}, \bibinfo {author} {\bibfnamefont {C.~L.}\ \bibnamefont {Hughlett}},
  \bibinfo {author} {\bibfnamefont {H.}~\bibnamefont {Zhou}}, \bibinfo {author}
  {\bibfnamefont {T.-H.}\ \bibnamefont {Chao}},\ and\ \bibinfo {author}
  {\bibfnamefont {J.~C.}\ \bibnamefont {Hanan}},\ }\bibfield  {title} {\bibinfo
  {title} {Neural network post-processing of grayscale optical correlator},\
  }in\ \href {https://doi.org/https://doi.org/10.1117/12.615573} {\emph
  {\bibinfo {booktitle} {Optical Information Systems III}}},\ Vol.\ \bibinfo
  {volume} {5908}\ (\bibinfo {organization} {SPIE},\ \bibinfo {year} {2005})\
  pp.\ \bibinfo {pages} {291--300}\BibitemShut {NoStop}%
\bibitem [{\citenamefont {Chang}\ \emph {et~al.}(2018)\citenamefont {Chang},
  \citenamefont {Sitzmann}, \citenamefont {Dun}, \citenamefont {Heidrich},\
  and\ \citenamefont {Wetzstein}}]{chang2018hybrid}%
  \BibitemOpen
  \bibfield  {author} {\bibinfo {author} {\bibfnamefont {J.}~\bibnamefont
  {Chang}}, \bibinfo {author} {\bibfnamefont {V.}~\bibnamefont {Sitzmann}},
  \bibinfo {author} {\bibfnamefont {X.}~\bibnamefont {Dun}}, \bibinfo {author}
  {\bibfnamefont {W.}~\bibnamefont {Heidrich}},\ and\ \bibinfo {author}
  {\bibfnamefont {G.}~\bibnamefont {Wetzstein}},\ }\bibfield  {title} {\bibinfo
  {title} {Hybrid optical-electronic convolutional neural networks with
  optimized diffractive optics for image classification},\ }\href
  {https://doi.org/https://doi.org/10.1038/s41598-018-30619-y} {\bibfield
  {journal} {\bibinfo  {journal} {Scientific reports}\ }\textbf {\bibinfo
  {volume} {8}},\ \bibinfo {pages} {12324} (\bibinfo {year}
  {2018})}\BibitemShut {NoStop}%
\bibitem [{\citenamefont {Miscuglio}\ \emph {et~al.}(2020)\citenamefont
  {Miscuglio}, \citenamefont {Hu}, \citenamefont {Li}, \citenamefont {George},
  \citenamefont {Capanna}, \citenamefont {Dalir}, \citenamefont {Bardet},
  \citenamefont {Gupta},\ and\ \citenamefont
  {Sorger}}]{miscuglio2020massively}%
  \BibitemOpen
  \bibfield  {author} {\bibinfo {author} {\bibfnamefont {M.}~\bibnamefont
  {Miscuglio}}, \bibinfo {author} {\bibfnamefont {Z.}~\bibnamefont {Hu}},
  \bibinfo {author} {\bibfnamefont {S.}~\bibnamefont {Li}}, \bibinfo {author}
  {\bibfnamefont {J.~K.}\ \bibnamefont {George}}, \bibinfo {author}
  {\bibfnamefont {R.}~\bibnamefont {Capanna}}, \bibinfo {author} {\bibfnamefont
  {H.}~\bibnamefont {Dalir}}, \bibinfo {author} {\bibfnamefont {P.~M.}\
  \bibnamefont {Bardet}}, \bibinfo {author} {\bibfnamefont {P.}~\bibnamefont
  {Gupta}},\ and\ \bibinfo {author} {\bibfnamefont {V.~J.}\ \bibnamefont
  {Sorger}},\ }\bibfield  {title} {\bibinfo {title} {Massively parallel
  amplitude-only fourier neural network},\ }\href
  {https://doi.org/https://doi.org/10.1364/OPTICA.408659} {\bibfield  {journal}
  {\bibinfo  {journal} {Optica}\ }\textbf {\bibinfo {volume} {7}},\ \bibinfo
  {pages} {1812} (\bibinfo {year} {2020})}\BibitemShut {NoStop}%
\bibitem [{\citenamefont {Goodman}(2005)}]{goodman2005introduction}%
  \BibitemOpen
  \bibfield  {author} {\bibinfo {author} {\bibfnamefont {J.~W.}\ \bibnamefont
  {Goodman}},\ }\href@noop {} {\emph {\bibinfo {title} {Introduction to Fourier
  optics}}}\ (\bibinfo  {publisher} {Roberts and Company publishers},\ \bibinfo
  {year} {2005})\BibitemShut {NoStop}%
\bibitem [{\citenamefont {Feldmann}\ \emph {et~al.}(2021)\citenamefont
  {Feldmann}, \citenamefont {Youngblood}, \citenamefont {Karpov}, \citenamefont
  {Gehring}, \citenamefont {Li}, \citenamefont {Stappers}, \citenamefont
  {Le~Gallo}, \citenamefont {Fu}, \citenamefont {Lukashchuk}, \citenamefont
  {Raja} \emph {et~al.}}]{feldmann2021parallel}%
  \BibitemOpen
  \bibfield  {author} {\bibinfo {author} {\bibfnamefont {J.}~\bibnamefont
  {Feldmann}}, \bibinfo {author} {\bibfnamefont {N.}~\bibnamefont
  {Youngblood}}, \bibinfo {author} {\bibfnamefont {M.}~\bibnamefont {Karpov}},
  \bibinfo {author} {\bibfnamefont {H.}~\bibnamefont {Gehring}}, \bibinfo
  {author} {\bibfnamefont {X.}~\bibnamefont {Li}}, \bibinfo {author}
  {\bibfnamefont {M.}~\bibnamefont {Stappers}}, \bibinfo {author}
  {\bibfnamefont {M.}~\bibnamefont {Le~Gallo}}, \bibinfo {author}
  {\bibfnamefont {X.}~\bibnamefont {Fu}}, \bibinfo {author} {\bibfnamefont
  {A.}~\bibnamefont {Lukashchuk}}, \bibinfo {author} {\bibfnamefont {A.~S.}\
  \bibnamefont {Raja}}, \emph {et~al.},\ }\bibfield  {title} {\bibinfo {title}
  {Parallel convolutional processing using an integrated photonic tensor
  core},\ }\href {https://doi.org/https://doi.org/10.1038/s41586-020-03070-1}
  {\bibfield  {journal} {\bibinfo  {journal} {Nature}\ }\textbf {\bibinfo
  {volume} {589}},\ \bibinfo {pages} {52} (\bibinfo {year} {2021})}\BibitemShut
  {NoStop}%
\bibitem [{\citenamefont {Mehrabian}\ \emph {et~al.}(2019)\citenamefont
  {Mehrabian}, \citenamefont {Miscuglio}, \citenamefont {Alkabani},
  \citenamefont {Sorger},\ and\ \citenamefont
  {El-Ghazawi}}]{mehrabian2019winograd}%
  \BibitemOpen
  \bibfield  {author} {\bibinfo {author} {\bibfnamefont {A.}~\bibnamefont
  {Mehrabian}}, \bibinfo {author} {\bibfnamefont {M.}~\bibnamefont
  {Miscuglio}}, \bibinfo {author} {\bibfnamefont {Y.}~\bibnamefont {Alkabani}},
  \bibinfo {author} {\bibfnamefont {V.~J.}\ \bibnamefont {Sorger}},\ and\
  \bibinfo {author} {\bibfnamefont {T.}~\bibnamefont {El-Ghazawi}},\ }\bibfield
   {title} {\bibinfo {title} {A winograd-based integrated photonics accelerator
  for convolutional neural networks},\ }\href
  {https://doi.org/https://doi.org/10.1109/JSTQE.2019.2957443} {\bibfield
  {journal} {\bibinfo  {journal} {IEEE Journal of Selected Topics in Quantum
  Electronics}\ }\textbf {\bibinfo {volume} {26}},\ \bibinfo {pages} {1}
  (\bibinfo {year} {2019})}\BibitemShut {NoStop}%
\bibitem [{\citenamefont {Zhu}\ \emph {et~al.}(2022)\citenamefont {Zhu},
  \citenamefont {Zou}, \citenamefont {Zhang}, \citenamefont {Shi},
  \citenamefont {Luo}, \citenamefont {Wang}, \citenamefont {Cai}, \citenamefont
  {Wan}, \citenamefont {Wang}, \citenamefont {Jiang} \emph
  {et~al.}}]{zhu2022space}%
  \BibitemOpen
  \bibfield  {author} {\bibinfo {author} {\bibfnamefont {H.}~\bibnamefont
  {Zhu}}, \bibinfo {author} {\bibfnamefont {J.}~\bibnamefont {Zou}}, \bibinfo
  {author} {\bibfnamefont {H.}~\bibnamefont {Zhang}}, \bibinfo {author}
  {\bibfnamefont {Y.}~\bibnamefont {Shi}}, \bibinfo {author} {\bibfnamefont
  {S.}~\bibnamefont {Luo}}, \bibinfo {author} {\bibfnamefont {N.}~\bibnamefont
  {Wang}}, \bibinfo {author} {\bibfnamefont {H.}~\bibnamefont {Cai}}, \bibinfo
  {author} {\bibfnamefont {L.}~\bibnamefont {Wan}}, \bibinfo {author}
  {\bibfnamefont {B.}~\bibnamefont {Wang}}, \bibinfo {author} {\bibfnamefont
  {X.}~\bibnamefont {Jiang}}, \emph {et~al.},\ }\bibfield  {title} {\bibinfo
  {title} {Space-efficient optical computing with an integrated chip
  diffractive neural network},\ }\href
  {https://doi.org/https://doi.org/10.1038/s41467-022-28702-0} {\bibfield
  {journal} {\bibinfo  {journal} {Nature communications}\ }\textbf {\bibinfo
  {volume} {13}},\ \bibinfo {pages} {1044} (\bibinfo {year}
  {2022})}\BibitemShut {NoStop}%
\bibitem [{\citenamefont {Colburn}\ \emph {et~al.}(2019)\citenamefont
  {Colburn}, \citenamefont {Chu}, \citenamefont {Shilzerman},\ and\
  \citenamefont {Majumdar}}]{colburn2019optical}%
  \BibitemOpen
  \bibfield  {author} {\bibinfo {author} {\bibfnamefont {S.}~\bibnamefont
  {Colburn}}, \bibinfo {author} {\bibfnamefont {Y.}~\bibnamefont {Chu}},
  \bibinfo {author} {\bibfnamefont {E.}~\bibnamefont {Shilzerman}},\ and\
  \bibinfo {author} {\bibfnamefont {A.}~\bibnamefont {Majumdar}},\ }\bibfield
  {title} {\bibinfo {title} {Optical frontend for a convolutional neural
  network},\ }\href {https://doi.org/https://doi.org/10.1364/AO.58.003179}
  {\bibfield  {journal} {\bibinfo  {journal} {Applied optics}\ }\textbf
  {\bibinfo {volume} {58}},\ \bibinfo {pages} {3179} (\bibinfo {year}
  {2019})}\BibitemShut {NoStop}%
\bibitem [{\citenamefont {Namias}(1980)}]{Nam80}%
  \BibitemOpen
  \bibfield  {author} {\bibinfo {author} {\bibfnamefont {V.}~\bibnamefont
  {Namias}},\ }\bibfield  {title} {\bibinfo {title} {The fractional order
  fourier transform and its application to quantum mechanics},\ }\href
  {https://doi.org/doi.org/10.1093/imamat/25.3.241} {\bibfield  {journal}
  {\bibinfo  {journal} {IMA Journal of Applied Mathematics}\ }\textbf {\bibinfo
  {volume} {25}},\ \bibinfo {pages} {241} (\bibinfo {year} {1980})}\BibitemShut
  {NoStop}%
\bibitem [{\citenamefont {Kutay}\ \emph {et~al.}(1997)\citenamefont {Kutay},
  \citenamefont {Ozaktas}, \citenamefont {Ankan},\ and\ \citenamefont
  {Onural}}]{Kut97}%
  \BibitemOpen
  \bibfield  {author} {\bibinfo {author} {\bibfnamefont {A.}~\bibnamefont
  {Kutay}}, \bibinfo {author} {\bibfnamefont {H.}~\bibnamefont {Ozaktas}},
  \bibinfo {author} {\bibfnamefont {O.}~\bibnamefont {Ankan}},\ and\ \bibinfo
  {author} {\bibfnamefont {L.}~\bibnamefont {Onural}},\ }\bibfield  {title}
  {\bibinfo {title} {Optimal filtering in fractional fourier domains},\ }\href
  {https://doi.org/10.1109/78.575688} {\bibfield  {journal} {\bibinfo
  {journal} {IEEE Transactions on Signal Processing}\ }\textbf {\bibinfo
  {volume} {45}},\ \bibinfo {pages} {1129} (\bibinfo {year}
  {1997})}\BibitemShut {NoStop}%
\bibitem [{\citenamefont {Ozturk}\ \emph {et~al.}(2021)\citenamefont {Ozturk},
  \citenamefont {Ozaktas}, \citenamefont {Gezici},\ and\ \citenamefont
  {Koç}}]{Ozt21}%
  \BibitemOpen
  \bibfield  {author} {\bibinfo {author} {\bibfnamefont {C.}~\bibnamefont
  {Ozturk}}, \bibinfo {author} {\bibfnamefont {H.~M.}\ \bibnamefont {Ozaktas}},
  \bibinfo {author} {\bibfnamefont {S.}~\bibnamefont {Gezici}},\ and\ \bibinfo
  {author} {\bibfnamefont {A.}~\bibnamefont {Koç}},\ }\bibfield  {title}
  {\bibinfo {title} {Optimal fractional fourier filtering for graph signals},\
  }\href {https://doi.org/10.1109/TSP.2021.3079804} {\bibfield  {journal}
  {\bibinfo  {journal} {IEEE Transactions on Signal Processing}\ }\textbf
  {\bibinfo {volume} {69}},\ \bibinfo {pages} {2902} (\bibinfo {year}
  {2021})}\BibitemShut {NoStop}%
\bibitem [{\citenamefont {Stankovi\'c}\ \emph {et~al.}(2003)\citenamefont
  {Stankovi\'c}, \citenamefont {Alieva},\ and\ \citenamefont
  {Bastiaans}}]{Sta03}%
  \BibitemOpen
  \bibfield  {author} {\bibinfo {author} {\bibfnamefont {L.}~\bibnamefont
  {Stankovi\'c}}, \bibinfo {author} {\bibfnamefont {T.}~\bibnamefont
  {Alieva}},\ and\ \bibinfo {author} {\bibfnamefont {M.~J.}\ \bibnamefont
  {Bastiaans}},\ }\bibfield  {title} {\bibinfo {title} {Time–frequency signal
  analysis based on the windowed fractional fourier transform},\ }\href
  {https://doi.org/https://doi.org/10.1016/S0165-1684(03)00197-X} {\bibfield
  {journal} {\bibinfo  {journal} {Signal Processing}\ }\textbf {\bibinfo
  {volume} {83}},\ \bibinfo {pages} {2459} (\bibinfo {year} {2003})},\ \bibinfo
  {note} {fractional Signal Processing and Applications}\BibitemShut {NoStop}%
\bibitem [{\citenamefont {Shi}\ \emph {et~al.}(2020)\citenamefont {Shi},
  \citenamefont {Zheng}, \citenamefont {Liu}, \citenamefont {Xiang},\ and\
  \citenamefont {Zhang}}]{Shi20}%
  \BibitemOpen
  \bibfield  {author} {\bibinfo {author} {\bibfnamefont {J.}~\bibnamefont
  {Shi}}, \bibinfo {author} {\bibfnamefont {J.}~\bibnamefont {Zheng}}, \bibinfo
  {author} {\bibfnamefont {X.}~\bibnamefont {Liu}}, \bibinfo {author}
  {\bibfnamefont {W.}~\bibnamefont {Xiang}},\ and\ \bibinfo {author}
  {\bibfnamefont {Q.}~\bibnamefont {Zhang}},\ }\bibfield  {title} {\bibinfo
  {title} {Novel short-time fractional fourier transform: Theory,
  implementation, and applications},\ }\href
  {https://doi.org/10.1109/TSP.2020.2992865} {\bibfield  {journal} {\bibinfo
  {journal} {IEEE Transactions on Signal Processing}\ }\textbf {\bibinfo
  {volume} {68}},\ \bibinfo {pages} {3280} (\bibinfo {year}
  {2020})}\BibitemShut {NoStop}%
\bibitem [{\citenamefont {Mendlovic}\ and\ \citenamefont
  {Ozaktas}(1993)}]{Men93}%
  \BibitemOpen
  \bibfield  {author} {\bibinfo {author} {\bibfnamefont {D.}~\bibnamefont
  {Mendlovic}}\ and\ \bibinfo {author} {\bibfnamefont {H.~M.}\ \bibnamefont
  {Ozaktas}},\ }\bibfield  {title} {\bibinfo {title} {Fractional fourier
  transforms and their optical implementation: I},\ }\href
  {https://doi.org/10.1364/JOSAA.10.001875} {\bibfield  {journal} {\bibinfo
  {journal} {J. Opt. Soc. Am. A}\ }\textbf {\bibinfo {volume} {10}},\ \bibinfo
  {pages} {1875} (\bibinfo {year} {1993})}\BibitemShut {NoStop}%
\bibitem [{\citenamefont {Almeida}(1997)}]{Alm97}%
  \BibitemOpen
  \bibfield  {author} {\bibinfo {author} {\bibfnamefont {L.}~\bibnamefont
  {Almeida}},\ }\bibfield  {title} {\bibinfo {title} {Product and convolution
  theorems for the fractional fourier transform},\ }\href
  {https://doi.org/10.1109/97.551689} {\bibfield  {journal} {\bibinfo
  {journal} {IEEE Signal Processing Letters}\ }\textbf {\bibinfo {volume}
  {4}},\ \bibinfo {pages} {15} (\bibinfo {year} {1997})}\BibitemShut {NoStop}%
\bibitem [{\citenamefont {Zayed}(1998)}]{Zay98}%
  \BibitemOpen
  \bibfield  {author} {\bibinfo {author} {\bibfnamefont {A.~I.}\ \bibnamefont
  {Zayed}},\ }\bibfield  {title} {\bibinfo {title} {A convolution and product
  theorem for the fractional fourier transform},\ }\href
  {https://doi.org/doi.org/10.1109/97.664179} {\bibfield  {journal} {\bibinfo
  {journal} {IEEE Signal Processing Letters}\ }\textbf {\bibinfo {volume}
  {5}},\ \bibinfo {pages} {101} (\bibinfo {year} {1998})}\BibitemShut {NoStop}%
\bibitem [{\citenamefont {Torres}\ \emph {et~al.}(2010)\citenamefont {Torres},
  \citenamefont {Pellat-Finet},\ and\ \citenamefont {Torres}}]{Tor10}%
  \BibitemOpen
  \bibfield  {author} {\bibinfo {author} {\bibfnamefont {R.}~\bibnamefont
  {Torres}}, \bibinfo {author} {\bibfnamefont {P.}~\bibnamefont
  {Pellat-Finet}},\ and\ \bibinfo {author} {\bibfnamefont {Y.}~\bibnamefont
  {Torres}},\ }\bibfield  {title} {\bibinfo {title} {Fractional convolution,
  fractional correlation and their translation invariance properties},\ }\href
  {https://doi.org/doi.org/10.1016/j.sigpro.2009.12.016} {\bibfield  {journal}
  {\bibinfo  {journal} {Signal Processing}\ }\textbf {\bibinfo {volume} {90}},\
  \bibinfo {pages} {1976} (\bibinfo {year} {2010})}\BibitemShut {NoStop}%
\bibitem [{\citenamefont {Shi}\ \emph {et~al.}(2014)\citenamefont {Shi},
  \citenamefont {Sha}, \citenamefont {Song},\ and\ \citenamefont
  {Zhang}}]{Shi14}%
  \BibitemOpen
  \bibfield  {author} {\bibinfo {author} {\bibfnamefont {J.}~\bibnamefont
  {Shi}}, \bibinfo {author} {\bibfnamefont {X.}~\bibnamefont {Sha}}, \bibinfo
  {author} {\bibfnamefont {X.}~\bibnamefont {Song}},\ and\ \bibinfo {author}
  {\bibfnamefont {N.}~\bibnamefont {Zhang}},\ }\bibfield  {title} {\bibinfo
  {title} {Generalized convolution theorem associated with fractional fourier
  transform},\ }\href {https://doi.org/doi.org/10.1002/wcm.2254} {\bibfield
  {journal} {\bibinfo  {journal} {Wireless Communications and Mobile
  Computing}\ }\textbf {\bibinfo {volume} {14}},\ \bibinfo {pages} {1340}
  (\bibinfo {year} {2014})}\BibitemShut {NoStop}%
\bibitem [{\citenamefont {Atakishiyev}\ and\ \citenamefont
  {Wolf}(1997)}]{Ata97}%
  \BibitemOpen
  \bibfield  {author} {\bibinfo {author} {\bibfnamefont {N.~M.}\ \bibnamefont
  {Atakishiyev}}\ and\ \bibinfo {author} {\bibfnamefont {K.~B.}\ \bibnamefont
  {Wolf}},\ }\bibfield  {title} {\bibinfo {title} {Fractional fourier--kravchuk
  transform},\ }\href {https://doi.org/10.1364/JOSAA.14.001467} {\bibfield
  {journal} {\bibinfo  {journal} {J. Opt. Soc. Am. A}\ }\textbf {\bibinfo
  {volume} {14}},\ \bibinfo {pages} {1467} (\bibinfo {year}
  {1997})}\BibitemShut {NoStop}%
\bibitem [{\citenamefont {Candan}\ \emph {et~al.}(2000)\citenamefont {Candan},
  \citenamefont {Kutay},\ and\ \citenamefont {Ozaktas}}]{Can00}%
  \BibitemOpen
  \bibfield  {author} {\bibinfo {author} {\bibfnamefont {C.}~\bibnamefont
  {Candan}}, \bibinfo {author} {\bibfnamefont {M.~A.}\ \bibnamefont {Kutay}},\
  and\ \bibinfo {author} {\bibfnamefont {H.~M.}\ \bibnamefont {Ozaktas}},\
  }\bibfield  {title} {\bibinfo {title} {The discrete fractional fourier
  transform},\ }\href@noop {} {\bibfield  {journal} {\bibinfo  {journal} {IEEE
  Transactions on signal processing}\ }\textbf {\bibinfo {volume} {48}},\
  \bibinfo {pages} {1329} (\bibinfo {year} {2000})}\BibitemShut {NoStop}%
\bibitem [{\citenamefont {Weimann}\ \emph {et~al.}(2016)\citenamefont
  {Weimann}, \citenamefont {Perez-Leija}, \citenamefont {Lebugle},
  \citenamefont {Keil}, \citenamefont {Tichy}, \citenamefont {Gr{\"a}fe},
  \citenamefont {Heilmann}, \citenamefont {Nolte}, \citenamefont {Moya-Cessa},
  \citenamefont {Weihs}, \citenamefont {Christodoulides},\ and\ \citenamefont
  {Szameit}}]{Wei16}%
  \BibitemOpen
  \bibfield  {author} {\bibinfo {author} {\bibfnamefont {S.}~\bibnamefont
  {Weimann}}, \bibinfo {author} {\bibfnamefont {A.}~\bibnamefont
  {Perez-Leija}}, \bibinfo {author} {\bibfnamefont {M.}~\bibnamefont
  {Lebugle}}, \bibinfo {author} {\bibfnamefont {R.}~\bibnamefont {Keil}},
  \bibinfo {author} {\bibfnamefont {M.}~\bibnamefont {Tichy}}, \bibinfo
  {author} {\bibfnamefont {M.}~\bibnamefont {Gr{\"a}fe}}, \bibinfo {author}
  {\bibfnamefont {R.}~\bibnamefont {Heilmann}}, \bibinfo {author}
  {\bibfnamefont {S.}~\bibnamefont {Nolte}}, \bibinfo {author} {\bibfnamefont
  {H.}~\bibnamefont {Moya-Cessa}}, \bibinfo {author} {\bibfnamefont
  {G.}~\bibnamefont {Weihs}}, \bibinfo {author} {\bibfnamefont {D.~N.}\
  \bibnamefont {Christodoulides}},\ and\ \bibinfo {author} {\bibfnamefont
  {A.}~\bibnamefont {Szameit}},\ }\bibfield  {title} {\bibinfo {title}
  {Implementation of quantum and classical discrete fractional fourier
  transforms},\ }\href {https://doi.org/10.1038/ncomms11027} {\bibfield
  {journal} {\bibinfo  {journal} {Nature Communications}\ }\textbf {\bibinfo
  {volume} {7}},\ \bibinfo {pages} {11027} (\bibinfo {year}
  {2016})}\BibitemShut {NoStop}%
\bibitem [{\citenamefont {Honari-Latifpour}\ \emph {et~al.}(2022)\citenamefont
  {Honari-Latifpour}, \citenamefont {Binaie}, \citenamefont {Eftekhar},
  \citenamefont {Madamopoulos},\ and\ \citenamefont {Miri}}]{Hon22}%
  \BibitemOpen
  \bibfield  {author} {\bibinfo {author} {\bibfnamefont {M.}~\bibnamefont
  {Honari-Latifpour}}, \bibinfo {author} {\bibfnamefont {A.}~\bibnamefont
  {Binaie}}, \bibinfo {author} {\bibfnamefont {M.~A.}\ \bibnamefont
  {Eftekhar}}, \bibinfo {author} {\bibfnamefont {N.}~\bibnamefont
  {Madamopoulos}},\ and\ \bibinfo {author} {\bibfnamefont {M.-A.}\ \bibnamefont
  {Miri}},\ }\bibfield  {title} {\bibinfo {title} {Arrayed waveguide lens for
  beam steering},\ }\href {https://doi.org/doi:10.1515/nanoph-2022-0198}
  {\bibfield  {journal} {\bibinfo  {journal} {Nanophotonics}\ }\textbf
  {\bibinfo {volume} {11}},\ \bibinfo {pages} {3679} (\bibinfo {year}
  {2022})}\BibitemShut {NoStop}%
\bibitem [{\citenamefont {Keshavarz}\ \emph {et~al.}(2023)\citenamefont
  {Keshavarz}, \citenamefont {Shariati},\ and\ \citenamefont
  {Miri}}]{keshavarz2023real}%
  \BibitemOpen
  \bibfield  {author} {\bibinfo {author} {\bibfnamefont {R.}~\bibnamefont
  {Keshavarz}}, \bibinfo {author} {\bibfnamefont {N.}~\bibnamefont
  {Shariati}},\ and\ \bibinfo {author} {\bibfnamefont {M.-A.}\ \bibnamefont
  {Miri}},\ }\bibfield  {title} {\bibinfo {title} {Real-time discrete
  fractional fourier transform using metamaterial coupled lines network},\
  }\href {https://doi.org/10.1109/TMTT.2023.3278929} {\bibfield  {journal}
  {\bibinfo  {journal} {IEEE Transactions on Microwave Theory and Techniques}\
  }\textbf {\bibinfo {volume} {71}},\ \bibinfo {pages} {3414} (\bibinfo {year}
  {2023})}\BibitemShut {NoStop}%
\bibitem [{\citenamefont {Tschernig}\ \emph {et~al.}(2018)\citenamefont
  {Tschernig}, \citenamefont {de~J.~Le\'{o}n-Montiel}, \citenamefont {{n}a
  Loaiza}, \citenamefont {Szameit}, \citenamefont {Busch},\ and\ \citenamefont
  {Perez-Leija}}]{Tsc18}%
  \BibitemOpen
  \bibfield  {author} {\bibinfo {author} {\bibfnamefont {K.}~\bibnamefont
  {Tschernig}}, \bibinfo {author} {\bibfnamefont {R.}~\bibnamefont
  {de~J.~Le\'{o}n-Montiel}}, \bibinfo {author} {\bibfnamefont {O.~S.~M.}\
  \bibnamefont {{n}a Loaiza}}, \bibinfo {author} {\bibfnamefont
  {A.}~\bibnamefont {Szameit}}, \bibinfo {author} {\bibfnamefont
  {K.}~\bibnamefont {Busch}},\ and\ \bibinfo {author} {\bibfnamefont
  {A.}~\bibnamefont {Perez-Leija}},\ }\bibfield  {title} {\bibinfo {title}
  {Multiphoton discrete fractional fourier dynamics in waveguide beam
  splitters},\ }\href {https://doi.org/10.1364/JOSAB.35.001985} {\bibfield
  {journal} {\bibinfo  {journal} {J. Opt. Soc. Am. B}\ }\textbf {\bibinfo
  {volume} {35}},\ \bibinfo {pages} {1985} (\bibinfo {year}
  {2018})}\BibitemShut {NoStop}%
\bibitem [{\citenamefont {Narducci}\ and\ \citenamefont
  {Orszag}(1972)}]{Nar72}%
  \BibitemOpen
  \bibfield  {author} {\bibinfo {author} {\bibfnamefont {L.~M.}\ \bibnamefont
  {Narducci}}\ and\ \bibinfo {author} {\bibfnamefont {M.}~\bibnamefont
  {Orszag}},\ }\bibfield  {title} {\bibinfo {title} {Eigenvalues and
  eigenvectors of angular momentum operator jx without the theory of
  rotations},\ }\href {https://doi.org/10.1119/1.1987068} {\bibfield  {journal}
  {\bibinfo  {journal} {American Journal of Physics}\ }\textbf {\bibinfo
  {volume} {40}},\ \bibinfo {pages} {1811} (\bibinfo {year} {1972})},\ \Eprint
  {https://arxiv.org/abs/https://doi.org/10.1119/1.1987068}
  {https://doi.org/10.1119/1.1987068} \BibitemShut {NoStop}%
\bibitem [{\citenamefont {Olver}\ \emph {et~al.}(2010)\citenamefont {Olver},
  \citenamefont {Lozier}, \citenamefont {Boisvert},\ and\ \citenamefont
  {Clark}}]{Olv10}%
  \BibitemOpen
  \bibfield  {author} {\bibinfo {author} {\bibfnamefont {F.}~\bibnamefont
  {Olver}}, \bibinfo {author} {\bibfnamefont {D.}~\bibnamefont {Lozier}},
  \bibinfo {author} {\bibfnamefont {R.}~\bibnamefont {Boisvert}},\ and\
  \bibinfo {author} {\bibfnamefont {C.}~\bibnamefont {Clark}},\ }\href@noop {}
  {\emph {\bibinfo {title} {NIST Handbook of Mathematical Functions}}}\
  (\bibinfo  {publisher} {Cambridge University Press},\ \bibinfo {address} {New
  York},\ \bibinfo {year} {2010})\BibitemShut {NoStop}%
\bibitem [{\citenamefont {Huang}(1994)}]{Huang94}%
  \BibitemOpen
  \bibfield  {author} {\bibinfo {author} {\bibfnamefont {W.-P.}\ \bibnamefont
  {Huang}},\ }\bibfield  {title} {\bibinfo {title} {Coupled-mode theory for
  optical waveguides: an overview},\ }\href
  {https://doi.org/10.1364/JOSAA.11.000963} {\bibfield  {journal} {\bibinfo
  {journal} {JOSA A}\ }\textbf {\bibinfo {volume} {11}},\ \bibinfo {pages}
  {963} (\bibinfo {year} {1994})}\BibitemShut {NoStop}%
\bibitem [{\citenamefont {Christodoulides}\ \emph {et~al.}(2003)\citenamefont
  {Christodoulides}, \citenamefont {Lederer},\ and\ \citenamefont
  {Silberberg}}]{Chr03}%
  \BibitemOpen
  \bibfield  {author} {\bibinfo {author} {\bibfnamefont {D.~N.}\ \bibnamefont
  {Christodoulides}}, \bibinfo {author} {\bibfnamefont {F.}~\bibnamefont
  {Lederer}},\ and\ \bibinfo {author} {\bibfnamefont {Y.}~\bibnamefont
  {Silberberg}},\ }\bibfield  {title} {\bibinfo {title} {Discretizing light
  behaviour in linear and nonlinear waveguide lattices},\ }\href
  {https://doi.org/10.1038/nature01936} {\bibfield  {journal} {\bibinfo
  {journal} {Nature}\ }\textbf {\bibinfo {volume} {424}},\ \bibinfo {pages}
  {817} (\bibinfo {year} {2003})}\BibitemShut {NoStop}%
\bibitem [{\citenamefont {Wei}\ and\ \citenamefont {Ran}(2013)}]{Wei13}%
  \BibitemOpen
  \bibfield  {author} {\bibinfo {author} {\bibfnamefont {D.}~\bibnamefont
  {Wei}}\ and\ \bibinfo {author} {\bibfnamefont {Q.}~\bibnamefont {Ran}},\
  }\bibfield  {title} {\bibinfo {title} {Multiplicative filtering in the
  fractional fourier domain},\ }\href
  {https://doi.org/10.1007/s11760-011-0261-5} {\bibfield  {journal} {\bibinfo
  {journal} {Signal, Image and Video Processing}\ }\textbf {\bibinfo {volume}
  {7}},\ \bibinfo {pages} {575} (\bibinfo {year} {2013})}\BibitemShut {NoStop}%
\bibitem [{Note1()}]{Note1}%
  \BibitemOpen
  \bibinfo {note} {This depends on the definition of the DFrFT operator, and
  the cyclic index identity $\alpha =4$ is also used in the
  literature.}\BibitemShut {Stop}%
\bibitem [{\citenamefont {Arnold~F.}\ and\ \citenamefont
  {Vasilii~B.}(1988)}]{Nik88}%
  \BibitemOpen
  \bibfield  {author} {\bibinfo {author} {\bibfnamefont {N.}~\bibnamefont
  {Arnold~F.}}\ and\ \bibinfo {author} {\bibfnamefont {U.}~\bibnamefont
  {Vasilii~B.}},\ }\href@noop {} {\emph {\bibinfo {title} {Special Functions of
  Mathematical Physics}}}\ (\bibinfo  {publisher} {Birkhäuser},\ \bibinfo
  {address} {Boston},\ \bibinfo {year} {1988})\BibitemShut {NoStop}%
\bibitem [{Note2()}]{Note2}%
  \BibitemOpen
  \bibinfo {note} {Oscillations in the discrete case are defined by their zero
  crossings. That is, a discrete function $\protect \bm {f}$ has a
  zero-crossing at $n$ if $f_{n}f_{n-1}<=0$.}\BibitemShut {Stop}%
\bibitem [{Note3()}]{Note3}%
  \BibitemOpen
  \bibinfo {note} {For maximum pooling, the $N=2m$-dimensional output is
  decomposed into $m$ sub-groups, each of dimension two, where only the maximum
  number within each sub-group is preserved.}\BibitemShut {Stop}%
\bibitem [{\citenamefont {Canny}(1986)}]{Can86}%
  \BibitemOpen
  \bibfield  {author} {\bibinfo {author} {\bibfnamefont {J.}~\bibnamefont
  {Canny}},\ }\bibfield  {title} {\bibinfo {title} {A computational approach to
  edge detection},\ }\href {https://doi.org/10.1109/TPAMI.1986.4767851}
  {\bibfield  {journal} {\bibinfo  {journal} {IEEE Transactions on Pattern
  Analysis and Machine Intelligence}\ }\textbf {\bibinfo {volume} {PAMI-8}},\
  \bibinfo {pages} {679} (\bibinfo {year} {1986})}\BibitemShut {NoStop}%
\bibitem [{\citenamefont {Woods}(2012)}]{Laplace}%
  \BibitemOpen
  \bibfield  {author} {\bibinfo {author} {\bibfnamefont {J.~W.}\ \bibnamefont
  {Woods}},\ }\href
  {https://doi.org/https://doi.org/10.1016/B978-0-12-381420-3.00007-2} {\emph
  {\bibinfo {title} {Image Enhancement and Analysis}}},\ \bibinfo {edition}
  {second edition}\ ed.\ (\bibinfo  {publisher} {Academic Press},\ \bibinfo
  {address} {Boston},\ \bibinfo {year} {2012})\ pp.\ \bibinfo {pages}
  {223--256}\BibitemShut {NoStop}%
\bibitem [{\citenamefont {Liu}\ \emph {et~al.}(2022)\citenamefont {Liu},
  \citenamefont {Feng}, \citenamefont {Tian}, \citenamefont {Zhao},
  \citenamefont {Jin}, \citenamefont {Ouyang}, \citenamefont {Zhu},\ and\
  \citenamefont {Guo}}]{liu2022thermo}%
  \BibitemOpen
  \bibfield  {author} {\bibinfo {author} {\bibfnamefont {S.}~\bibnamefont
  {Liu}}, \bibinfo {author} {\bibfnamefont {J.}~\bibnamefont {Feng}}, \bibinfo
  {author} {\bibfnamefont {Y.}~\bibnamefont {Tian}}, \bibinfo {author}
  {\bibfnamefont {H.}~\bibnamefont {Zhao}}, \bibinfo {author} {\bibfnamefont
  {L.}~\bibnamefont {Jin}}, \bibinfo {author} {\bibfnamefont {B.}~\bibnamefont
  {Ouyang}}, \bibinfo {author} {\bibfnamefont {J.}~\bibnamefont {Zhu}},\ and\
  \bibinfo {author} {\bibfnamefont {J.}~\bibnamefont {Guo}},\ }\bibfield
  {title} {\bibinfo {title} {Thermo-optic phase shifters based on
  silicon-on-insulator platform: State-of-the-art and a review},\ }\href
  {https://doi.org/10.1007/s12200-022-00012-9} {\bibfield  {journal} {\bibinfo
  {journal} {Frontiers of Optoelectronics}\ }\textbf {\bibinfo {volume} {15}},\
  \bibinfo {pages} {9} (\bibinfo {year} {2022})}\BibitemShut {NoStop}%
\bibitem [{\citenamefont {Harris}\ \emph {et~al.}(2014)\citenamefont {Harris},
  \citenamefont {Ma}, \citenamefont {Mower}, \citenamefont {Baehr-Jones},
  \citenamefont {Englund}, \citenamefont {Hochberg},\ and\ \citenamefont
  {Galland}}]{harris2014efficient}%
  \BibitemOpen
  \bibfield  {author} {\bibinfo {author} {\bibfnamefont {N.~C.}\ \bibnamefont
  {Harris}}, \bibinfo {author} {\bibfnamefont {Y.}~\bibnamefont {Ma}}, \bibinfo
  {author} {\bibfnamefont {J.}~\bibnamefont {Mower}}, \bibinfo {author}
  {\bibfnamefont {T.}~\bibnamefont {Baehr-Jones}}, \bibinfo {author}
  {\bibfnamefont {D.}~\bibnamefont {Englund}}, \bibinfo {author} {\bibfnamefont
  {M.}~\bibnamefont {Hochberg}},\ and\ \bibinfo {author} {\bibfnamefont
  {C.}~\bibnamefont {Galland}},\ }\bibfield  {title} {\bibinfo {title}
  {Efficient, compact and low loss thermo-optic phase shifter in silicon},\
  }\href {https://doi.org/10.1364/OE.22.010487} {\bibfield  {journal} {\bibinfo
   {journal} {Optics express}\ }\textbf {\bibinfo {volume} {22}},\ \bibinfo
  {pages} {10487} (\bibinfo {year} {2014})}\BibitemShut {NoStop}%
\bibitem [{\citenamefont {R{\'\i}os}\ \emph {et~al.}(2022)\citenamefont
  {R{\'\i}os}, \citenamefont {Du}, \citenamefont {Zhang}, \citenamefont
  {Popescu}, \citenamefont {Shalaginov}, \citenamefont {Miller}, \citenamefont
  {Roberts}, \citenamefont {Kang}, \citenamefont {Richardson}, \citenamefont
  {Gu} \emph {et~al.}}]{rios2022ultra}%
  \BibitemOpen
  \bibfield  {author} {\bibinfo {author} {\bibfnamefont {C.}~\bibnamefont
  {R{\'\i}os}}, \bibinfo {author} {\bibfnamefont {Q.}~\bibnamefont {Du}},
  \bibinfo {author} {\bibfnamefont {Y.}~\bibnamefont {Zhang}}, \bibinfo
  {author} {\bibfnamefont {C.-C.}\ \bibnamefont {Popescu}}, \bibinfo {author}
  {\bibfnamefont {M.~Y.}\ \bibnamefont {Shalaginov}}, \bibinfo {author}
  {\bibfnamefont {P.}~\bibnamefont {Miller}}, \bibinfo {author} {\bibfnamefont
  {C.}~\bibnamefont {Roberts}}, \bibinfo {author} {\bibfnamefont
  {M.}~\bibnamefont {Kang}}, \bibinfo {author} {\bibfnamefont {K.~A.}\
  \bibnamefont {Richardson}}, \bibinfo {author} {\bibfnamefont
  {T.}~\bibnamefont {Gu}}, \emph {et~al.},\ }\bibfield  {title} {\bibinfo
  {title} {Ultra-compact nonvolatile phase shifter based on electrically
  reprogrammable transparent phase change materials},\ }\href
  {https://doi.org/10.1186/s43074-022-00070-4} {\bibfield  {journal} {\bibinfo
  {journal} {PhotoniX}\ }\textbf {\bibinfo {volume} {3}},\ \bibinfo {pages}
  {26} (\bibinfo {year} {2022})}\BibitemShut {NoStop}%
\bibitem [{\citenamefont {Dickinson}\ and\ \citenamefont
  {Steiglitz}(1982)}]{Dic82}%
  \BibitemOpen
  \bibfield  {author} {\bibinfo {author} {\bibfnamefont {B.}~\bibnamefont
  {Dickinson}}\ and\ \bibinfo {author} {\bibfnamefont {K.}~\bibnamefont
  {Steiglitz}},\ }\bibfield  {title} {\bibinfo {title} {Eigenvectors and
  functions of the discrete fourier transform},\ }\href
  {https://doi.org/10.1109/TASSP.1982.1163843} {\bibfield  {journal} {\bibinfo
  {journal} {IEEE Transactions on Acoustics, Speech, and Signal Processing}\
  }\textbf {\bibinfo {volume} {30}},\ \bibinfo {pages} {25} (\bibinfo {year}
  {1982})}\BibitemShut {NoStop}%
\bibitem [{\citenamefont {Bose}(2001)}]{Bos01}%
  \BibitemOpen
  \bibfield  {author} {\bibinfo {author} {\bibfnamefont {N.}~\bibnamefont
  {Bose}},\ }\bibfield  {title} {\bibinfo {title} {Eigenvectors and eigenvalues
  of 1-d and n-d dft matrices},\ }\href
  {https://doi.org/https://doi.org/10.1078/1434-8411-00019} {\bibfield
  {journal} {\bibinfo  {journal} {AEU - International Journal of Electronics
  and Communications}\ }\textbf {\bibinfo {volume} {55}},\ \bibinfo {pages}
  {131} (\bibinfo {year} {2001})}\BibitemShut {NoStop}%
\end{thebibliography}%

\end{document}